\documentclass[twocolumn]{aastex6}
\usepackage{natbib}
\usepackage{color}
\usepackage{soul}
\usepackage{graphicx}
\usepackage[T1]{fontenc}
\begin{document}
\title{Understanding formation of young, distributed low-mass stars and clusters in the W4 cloud complex }
\author{Neelam Panwar  \altaffilmark{1,3}, Manash R. Samal \altaffilmark{2}, A. K. Pandey \altaffilmark{1}, H. P. Singh \altaffilmark{3}, Saurabh Sharma \altaffilmark{1}}
\altaffiltext{1}{Aryabhatta Research Institute of Observational Sciences (ARIES), Nainital - 263001, India}
\altaffiltext{2}{Physical Research Laboratory, Navrangpura, Ahmedabad - 380009, India}
\altaffiltext{3}{Department of Physics \& Astrophysics, University of Delhi, Delhi-110007, India}
\begin{abstract}
It is well known that most of the stars form in rich clusters. However, recent $Spitzer$ observations have shown that a significant number of stars also form in distributed mode, origin of which is not well understood. In this work, we aim to investigate clustered and distributed mode of star formation in the W4 complex. To do so, we identified and characterized the 
young stellar population associated with the region using homogeneous infra-red data-sets obtained from 
2MASS, GLIMPSE, MIPS and WISE surveys. We make stellar surface density and minimum spanning tree maps to identify young clusters,
and use {\it Spitzer} images to identify irradiated structures, such as elephant trunk-like structures (ETLSs) 
and pillars in the region. 
The surface density distribution 
of the young stellar objects (YSOs) reveals three new clusterings and $\sim$~50\% distributed protostars in the H{\sc ii} region. The clusters are of low-mass nature 
but significantly younger than the central cluster IC~1805. 
We identified $\sim$~38 ETLSs in the region, a majority of which consist of one or a few stars 
at their tips. We find these stars are low-mass ($<$ 2~M$_\odot$) YSOs, located at the outskirts ($>$~17~pc) of the cluster IC~1805 
and are part of scattered distributed population. We argued that the star formation in the ETLSs of W4 is going on possibly due to 
triggering effect of expanding W4 bubble. Although high-resolution photometric and spectroscopic data would be required to confirm the scenario, nonetheless, we discuss the implications of this scenario for our understanding of distributed low-mass star
formation in cloud complexes as opposed to other mechanisms such as 
turbulent fragmentation and dynamical ejection. 

 \end{abstract}

\keywords{
stars : formation  - stars : pre-main-sequence - ISM : globules ­ H{\sc ii} regions - open cluster: initial mass function; star formation.
}
\section{Introduction}
Observations of many star forming regions (SFRs) have shown that cluster or group is the dominant mode of star formation.  
Stellar clustering results from the fractal properties of the molecular clouds under the effects of 
turbulence \citep{elm97,elm14}. 
Hence, star clusters can be used to investigate star formation from small scales (in small dense clouds) to the larger scales (in giant molecular clouds). 
At the same time, mid-infrared surveys such as {\it Spitzer} have shown that significant number 
of stars form in a more distributed mode. For example, \citet{eva09} studied several nearby molecular clouds and found that in each cloud 
a population of isolated young stellar objects (YSOs) is distributed throughout the cloud \citep[see also][]{koe08}. The origin of  this distributed population is unclear so far.
It is also not clear under what circumstances distributed mode of star formation would be more prevalent. 

In this work we aim to understand the formation of clusters and distributed mode of star formation by investigating 
young stellar content of the W4 complex.  
The W4 complex (see Fig. 1), located at a distance of $\sim$ 2 kpc in the Perseus arm of the Galaxy, 
is a part of the W3/W4/W5 cloud complex. It is 
ionized by the massive members of the cluster IC~1805 \citep{mas95}. The cluster IC~1805 is located at the central part
of the W4 H{\sc ii} region and embedded in a very low-extinction cloud of mean $E$($B - V$) $\sim$ 0.8 mag \citep[see]
[and references therein]{pan17}. It is a young cluster with an age of $\sim$ 2.5 Myr and total mass of 
$\sim$ 2700 M$_\odot$ \citep{pan17,sung17}. It harbors dozens of massive stars (O- or early B-type), 
of which the three most massive are BD +60502 (O4.5III), BD +60501 (O7V) and  HD 15570 (O4I) \citep{mas95,lef95},
hence more energetic than the Orion Nebula \citep[e.g., see discussion in][]{pan18}.
These massive 
members of the cluster IC~1805 have created a central cavity of $\sim$ 40 pc through mechanical 
and radiation output. 
In fact {\it Spitzer} and {\it Herschel} 
images of the H{\sc ii} region show that it has a bipolar bubble morphology \citep{pan17}, 
similar to that found by \citet[][]{deh15,sam18b}.
The ionizing sources have a considerable effect on the structure of the molecular
material as W4 bubble contains many bright-rimmed 
clouds (BRCs), namely, BRC~5, BRC~6, BRC~7, BRC~8, BRC~9 and a cometary globule CG~7S, 
which are thought to arise from the impact of UV photons of nearby massive stars on
pre-existing dense molecular material \citep{sug91,lef95,mor04,mia09,bis11}. Most of these BRCs contain Class~0/I sources at their heads which may be the
products of triggering due to radiation pressure from the massive members of the cluster IC~1805. The age distributions of 
the YSOs as a function of the distance from the most massive stars of the IC~1805)
 in BRCs 5 and 7 manifest age gradients \citep{pan14} supporting the model of triggered sequential star formation in these clouds.

This paper is organized as follows. We describe the datasets used in the present work in Sec. 2.
 Sec. 3 describes identification of candidate YSOs in the H{\sc ii} region. 
We discuss the incidence of clusterings and isolated distribution of YSOs in the region and also
their possible origin in Sec. 4. In Sec. 5, we provide conclusions of our study of the clustered and distributed mode of star formation in the complex and our main results.
\section{Data Used}
\begin{figure}
\centering
\includegraphics[scale = 0.45, trim =0  0 0 0, clip]{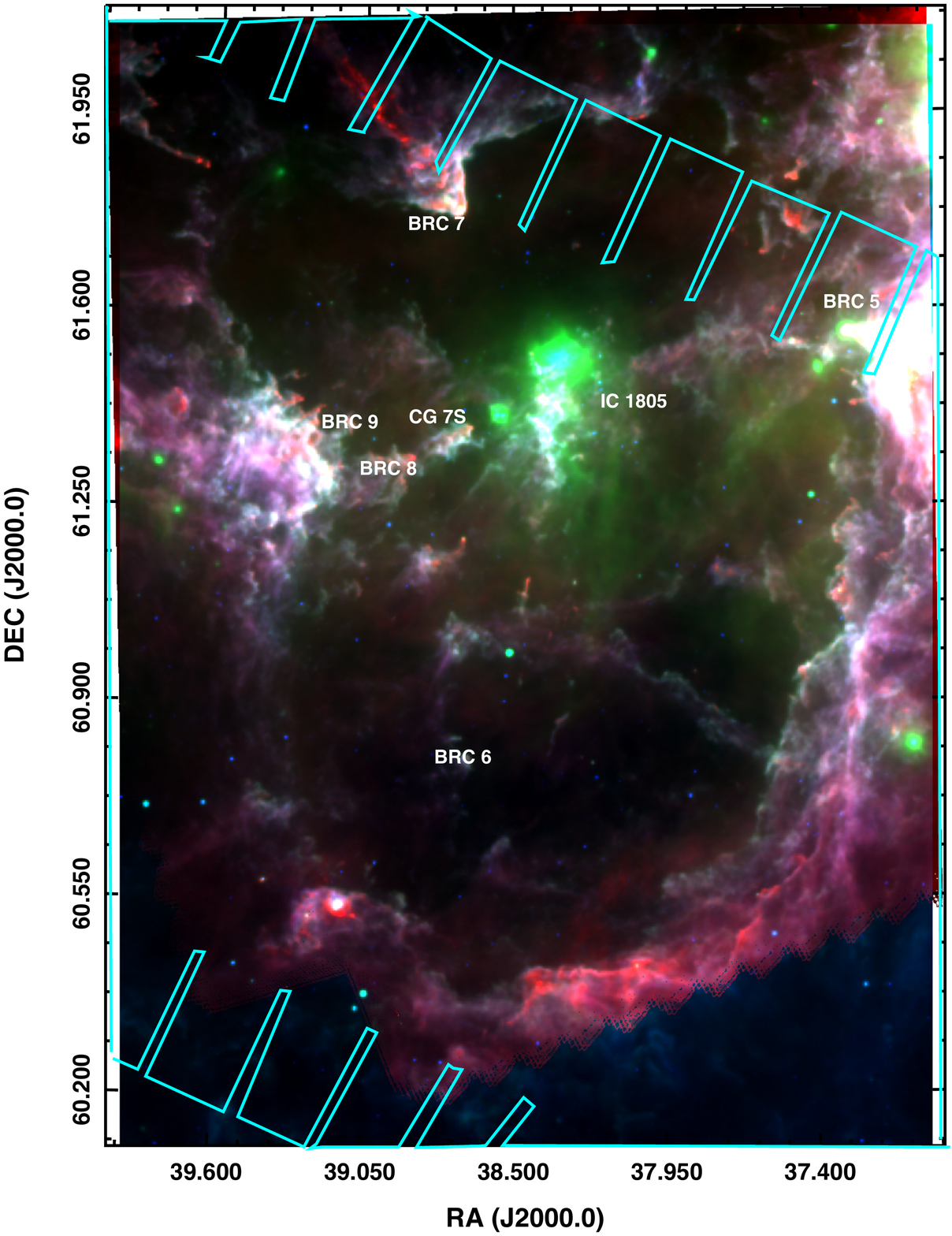}
\caption{Color-composite image (red: Herschel 500 $\micron$, green: WISE 22 $\micron$ and blue: WISE 12 $\micron$) of W4 H{\sc ii} region. The locations of central ionizing cluster IC~1805 alongwith BRCs and cometary globule are also marked. Cyan polygon represents the area covered by MIPS 24 $\micron$ observations.}
\end{figure}
The W4 bubble was observed in 24~$\mu$m using the Multi-band Imaging Photometer for 
{\it Spitzer} (MIPS) in two epochs. However, a part of the bubble (see upper-right corner of Fig. 1) 
is missing in these observations. 
We downloaded the MIPS post basic calibrated data images from the $Spitzer$ archive which were created at the image scale of 
2$^{\prime\prime}$.45 per pixel. The MOPEX-APEX pipeline was used to detect the point sources and to
perform the point response function (PRF) fitting photometry. All the sources were visually examined 
and any ambiguous sources were removed. In addition, we also included point
sources manually which were not automatically detected by APEX by supplying their co-ordinates to the Apex
 user-list module of MOPEX software to perform the PRF fitting and to extract the fluxes. The zero-point 
value of 7.14 Jy (adopted from MIPS Data Handbook) is used to convert the flux densities to magnitudes. 

We also used the archival infrared array camera (IRAC) data from GLIMPSE-360 (P.I. - B. A. 
Whitney). The GLIMPSE-360 observations were taken during the $Spitzer$ warm mission and hence were available only at 3.6 $\mu$m and 4.5 $\mu$m. 
To ensure the good quality data, we considered sources with photometric uncertainties $\leq$ 0.2 mag in each band. 

The Wide-field Infrared Survey Explorer (WISE) has explored the sky at four wavebands (3.4, 4.6, 12, 
and 22 $\mu$m). We have used the $WISE$ catalog from \citet{cut14}. We selected only sources that have 
contamination and confusion flags "000" in 3.4, 4.6 and 12 $\micron$ wavebands. 
NIR $JHK_s$ data for the stars in the H{\sc ii} region have been obtained from the Two
Micron All Sky Survey (2MASS) Point Source Catalog (PSC) \citep{cut03}. 

 For making various color-color diagrams to identify YSOs in the region, we band-merge all of the above catalogues as follows. We matched 
the GLIMPSE data with 2MASS catalog using a maximum 1$^{\prime\prime}$ radial tolerance \citep{gut08,pan14}. We integrated MIPS 24 $\micron$ catalog 
using a 2$^{\prime\prime}$.5 maximum radial tolerance with the GLIMPSE catalog \citep{meg12}. In  cases, where there was more than
one source within the matching radius, the closest
one is considered as the best match.

\section {Analysis : Identification of Young stellar Objects}
\label{sec:ana}
YSOs trace the most recent sites of star formation in a SFR as they are too young to undergo dynamical 
evolution and can be used to understand the star formation history and other properties of the region \citep{sni09,deh12,pan13,pan13a,bern16,yad16,saur17}. 
They exhibit excess emission in IR wavelengths due to their circumstellar disks. Hence, NIR to MIR photometric surveys 
of SFRs are useful tools to identify YSOs. YSOs occupy distinct 
positions in infrared color-color (CC) diagrams due to their distinct characteristics
which can be used for identification of young stellar
content of a complex. In this work, we are interested in tracing recent star formation activity of the complex. 
Therefore, we only searched for Class I and Class II YSOs of the complex. We describe several 
color diagnosis to identify such YSOs of the W4 complex.

\begin{figure*}
\includegraphics[scale = 0.45, trim = 0 120 0 150, clip]{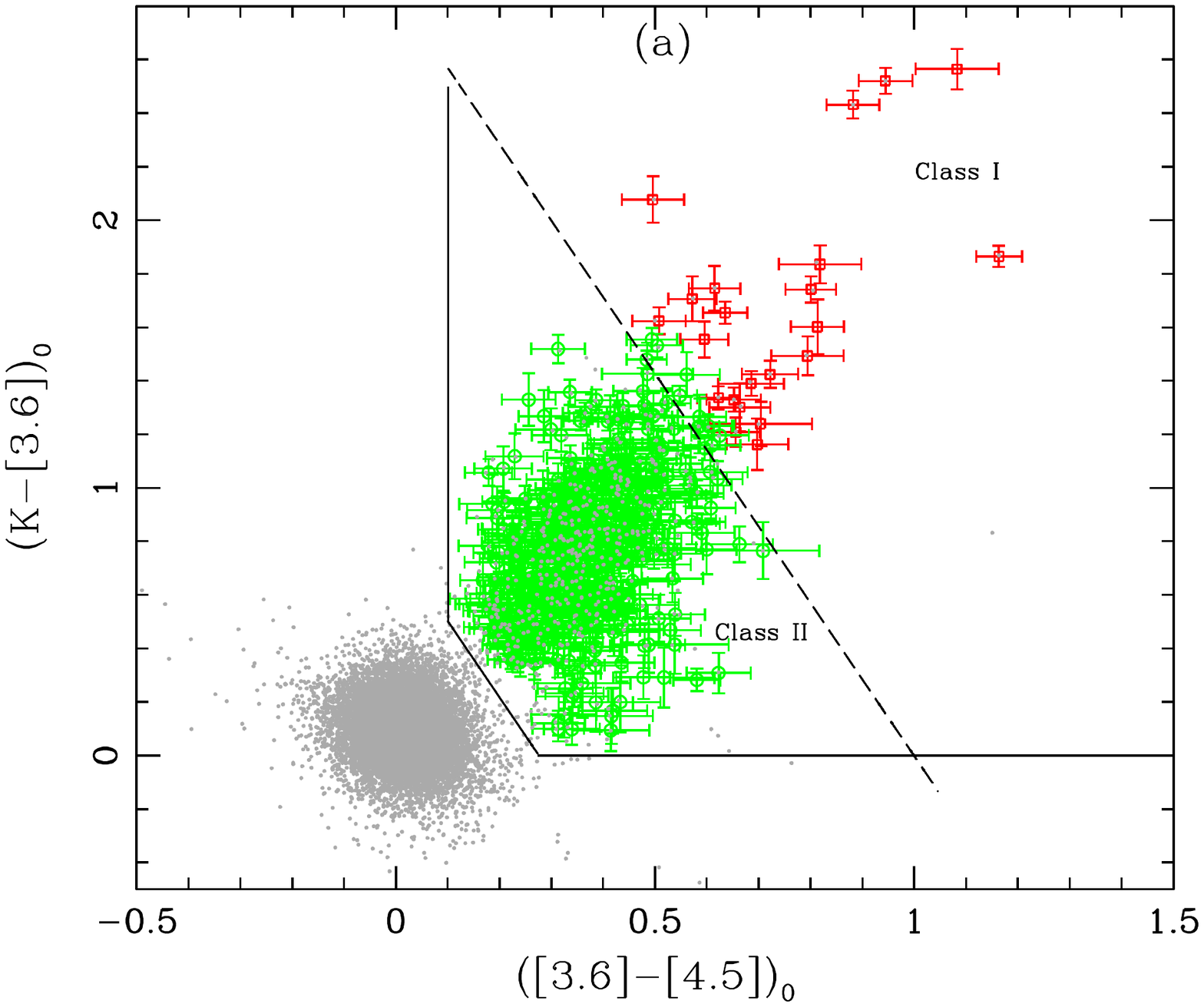}
\includegraphics[scale = 0.45, trim = 0 120 0 150, clip]{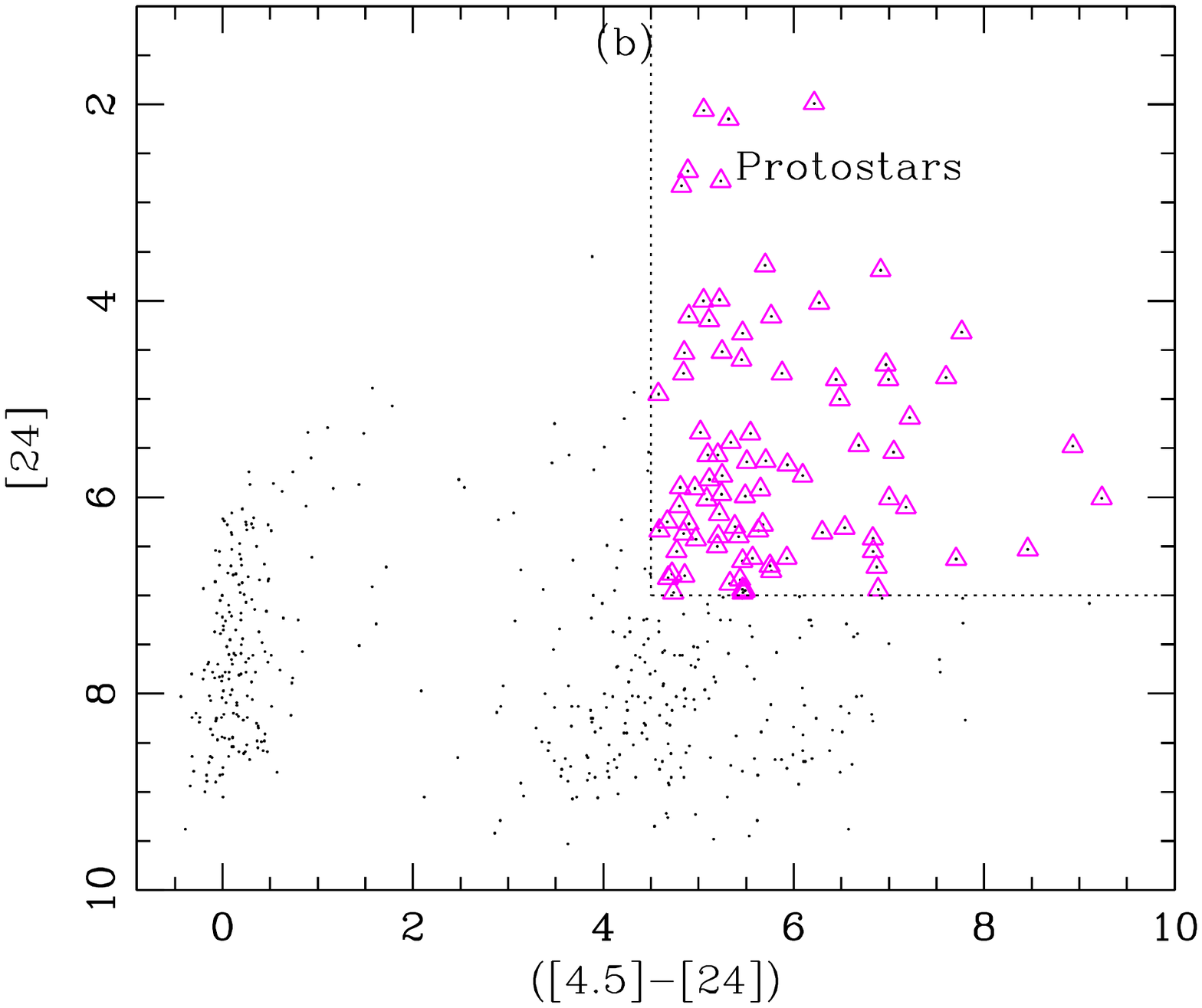}
\includegraphics[scale = 0.45, trim = 0 120 0 150, clip]{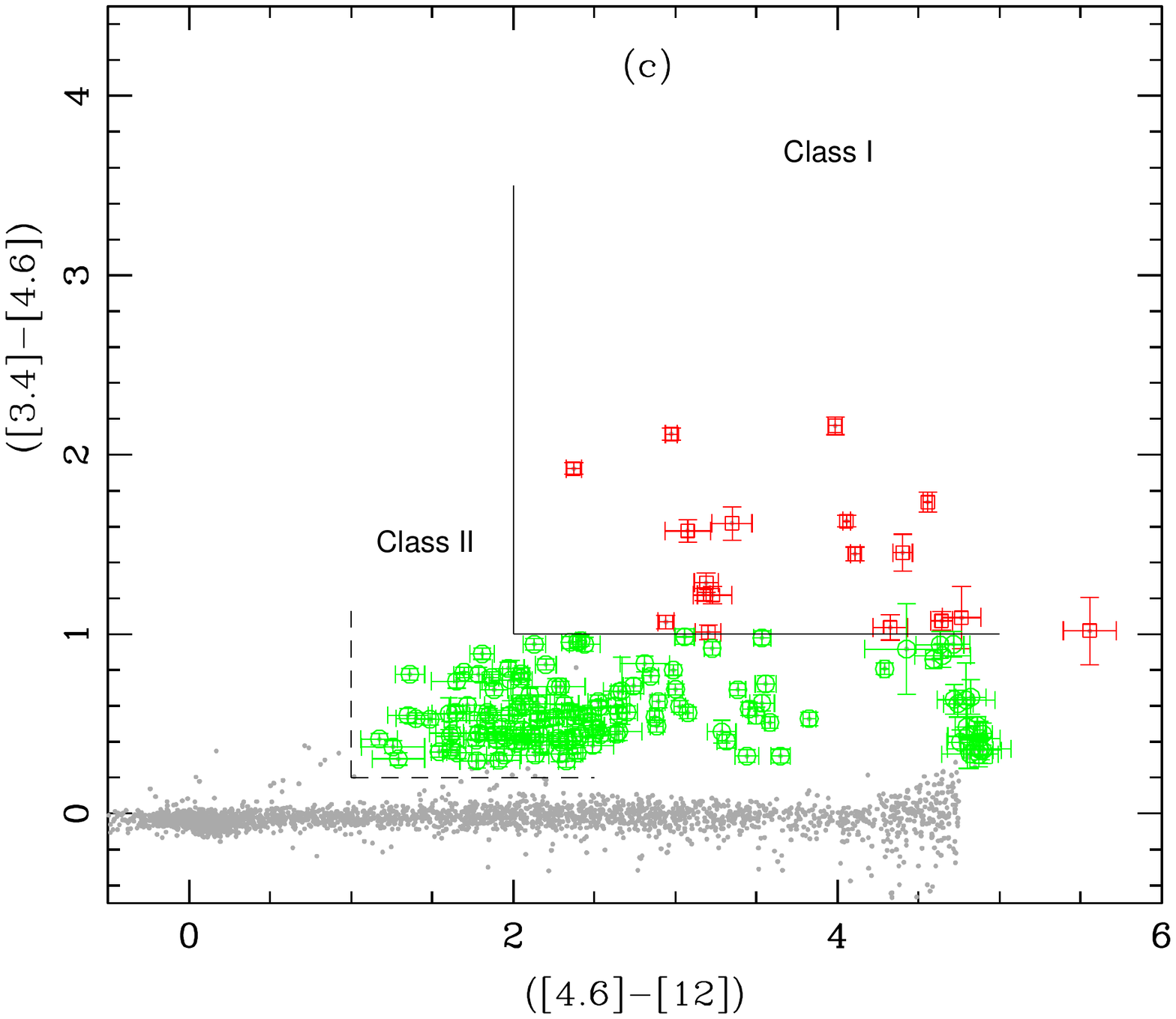}

\caption{(a) GLIMPSE/2MASS (K$_s$ - [3.6])$_0$ vs. ([3.6] - [4.5])$_0$ color-color diagram, (b) [24] vs. ([4.5] - [24]) color-magnitude diagram 
and (c) WISE ([3.4]-[4.6])/ ([4.6]-[12.0]) color-color 
diagram for the sources in W4 region. Red squares represent the Class~I and green circles represent Class~II sources. 
Magenta triangles represent the protostellar sources identified based on the MIPS/GLIMPSE color-magnitude criterion.
}
\end{figure*}
\subsection{YSOs using 2MASS/IRAC colors}
The IRAC CC diagrams are often used to identify and classify the YSO populations in SFRs 
\citep{meg04,all04}. In the absence of IRAC higher wavelength observations, IRAC 3.6, 4.5 $\micron$ data can be 
combined with 2MASS $H$ and $K_s$ data to identify YSOs of a complex \citep{gut08,gut09}. 
Following \citet{gut09}, we used IRAC/2MASS intrinsic K-[3.6] / [3.6]-[4.5] color-color diagram to identify the YSOs 
and classify them. In order to implement this method, we first generated the extinction map 
for the region using 2MASS data following the procedure described in \citet{pan14}.  
     In brief, we compute stellar extinction maps with the method proposed by \citet{cam02}. We divided the region into small bins and 
calculated the (H - K) colors of each bin as the median of the (H-K) colors of the 10 nearest neighbours. We then calculated 
E($H$ - $K$) = ($H$ - $K$) - $($H$ - $K$)_0$,
where $($H$ - $K$)_0$ is the median color of the stars in field region. We calculated $A_{K_s}$ 
for each bin using the relation $A_{K_s}$=1.82 $\times$ E(H - K). We then made an extinction map 
for the whole region.  We dereddened all those stars detected in K, 3.6 and 4.5 $\micron$ using the
extinction laws of \citet{fla07}. We used the $(K-[3.6])_0$ and $([3.6]-[4.5])_0$ color cuts to select Class I and Class II sources as described in \citet{gut09}. To minimize contamination from the extragalactic sources, especially PAH emitting galaxies and active galactic nuclei (AGN) in our YSO sample, we selected objects having [3.6] $<$ 14.5 mag \citep{gut09}.
 By this method we obtained 398 Class II and 23 Class I sources in the region. 
Fig. 2a shows the  $([3.6]-[4.5])_0$ vs. $(K-[3.6])_0$ CC diagram for the candidate YSOs. 
\subsection{Additional YSOs from 24 $\mu$m} 
There may be 
some sources which are deeply embedded and were not detected at any of the 2MASS/IRAC bands but possess MIPS photometry. To identify the additional YSOs and to examine the YSOs identified using 2MASS/IRAC colours, we included the sources which  have 24 $\mu$m counterparts. We found that out of $\sim$ 791 point sources with magnitude uncertainty $\le$ 0.2 mag within the region shown in Fig. 1, 
$\sim$ 550 MIPS sources have counterparts in GLIMPSE data. We considered sources which appear bright in MIPS 24 $\mu$m ([24] $<$ 7 mag) 
and have very red colors ([4.5]--[24] $>$ 4.5 mag) as Class~I sources \citep{gut09}. We choose brighter sources as at fainter magnitudes the extragalactic contamination may dominate our YSO sample \citep[e.g.,][]{meg09,gut09}. Using these criteria, we found $\sim$ 92 protostellar candidates, which are shown with magenta triangles
in MIPS/IRAC color-magnitude diagram (CMD) in Fig. 2b. 43 candidates out of these were identified using 2MASS/GLIMPSE data.  
All the previously identified protostars having 24 $\mu$m detections were checked and considered as Class~II 
sources if they did not have ([4.5] - [24]) $>$ 4 mag \citep{gut09}. Out of previously identified 23 Class I sources, 6 were reclassified as Class II sources. 
\subsection{YSOs from WISE data} 
We also used the WISE 3.4, 4.6 and 12 $\micron$ data for characterization of the embedded YSOs using the approach developed by
\citet{koe12}. This selection method uses a series of color and magnitude
 cuts to remove the contaminants such as background galaxies and nebulosity blobs. We identified 142 Class~II and 18 Class~I sources
 in the region. Out of these, 106 YSOs are already identified using aforementioned schemes. 
Fig. 2c shows the WISE CC diagram for the sources in the W4 H{\sc ii} region. Green Circles and red squares represent Class~I and Class~II sources, respectively.
\subsection{YSOs Summary and Completeness}

 Our final catalog includes the YSO candidates identified from
$Spitzer$ MIPS, GLIMPSE/2MASS and WISE data sets. We may point out that broad-line AGNs, unresolved knots of shock emission 
and faint sources contaminated by copious PAH nebulosity are likely to affect our YSO selection. Using the 
criteria of \citet{gut09,koe12}, we have removed these contaminants from our YSO catalog. Although 
these criteria for a distant region ($\sim$ 2 kpc) may lead to an overestimation of the contamination, 
this would ensure the high reliability of our YSO catalog. \citet{rob08} found that in their sample of sources with redder IRAC/MIPS 
colors, approximately 30\%-50\% may be asymptotic giant branch (AGB) stars whereas planetary nebulae and background galaxies together may be
 at most 2\%-3\% of the total. AGB stars show a steep spectral index at long wavelengths. Based on their
SEDs, \citet{rob08} used a criteria [8.0] - [24] $<$ 2.2 mag
to identify them. In the absence of IRAC 8.0 $\micron$ data, we
are unable to apply \citet{rob08} criteria to eliminate such sources. However, \citet{rob08} approximated the surface density distribution of candidate AGB stars in different longitude and latitude bins by a simple function. This function suggests a negligible contribution from the AGB stars ($\sim$ 0.004) within the region 
covered in the present study. 
In total, we found $\sim$ 524 candidate YSOs in the direction of the W4 complex. Out of these $\sim$ 100 are Class I sources. A sample
list of the YSO candidates with their magnitudes in different bands
is given in Table 1 and the entire table is available
in electronic form only.

The peak of the observed luminosity function can be assumed to estimate the 90\% completeness limit \citep{eva03,jose16}. 
In order to evaluate the completeness of the MIPS, IRAC/2MASS and WISE data, we plot the histograms of the sources detected in each band. The completeness limits are 8.0 mag, 13.5 mag and 13.0 mag in 24 $\micron$, 3.6 $\micron$ and 4.5 $\micron$, 
respectively. The completeness limits of WISE data are estimated as 12.5, 12.5, 9.5 and 7.0 mag in the 3.4, 4.6, 12 and 22 $\micron$ wavebands, respectively. In our analysis, we have used only first three WISE wavelengths. Therefore, the completeness of the WISE sources is limited by the bright nebulosity and relatively low sensitivity at 12 $\micron$. 
In this work, majority of the YSOs have been identified using IRAC/2MASS CC diagram (see Fig. 2a). We found 90\% 
of the 3.6 $\micron$ and 4.5 $\micron$ sources have K$_s$-band counterparts. 
Therefore, majority of the IRAC sources have been detected in K$_s$-band. Thus our YSO completeness limit  is primarily limited by the completeness of K$_s$ band, which is $K_s$ $\sim$ 14.5 mag. 
 We note that the local completeness limits can be affected by 
the variable reddening and crowding of the sources across the region.
 Using the evolutionary models of \citet{sie00} for an age of 
2 Myr (lifetime of Class II phase of YSOs) and a distance of 2 kpc, the mass completeness corresponds to $\sim$ 0.5 M$_\odot$.

\begin{table*}
\centering
\caption{YSOs from MIPS, IRAC/2MASS and WISE data. The entire table is available in electronic form. }
\tiny
\hspace{-2.0cm}
\begin{tabular}{ccccccccccccc}
\hline
Id     & RA & DEC & J$\pm$eJ & H$\pm$eH & $K_s$$\pm$e$K_s$& [3.6]$\pm$ & [4.5]$\pm$ &[24]$\pm$ &[3.4]$\pm$&[4.6]$\pm$ &[12]$\pm$\\
       & (J2000) &(J2000)&   &          &         &  e[3.6]    &  e[4.5]    & e[24]    &   e[3.4] & e[4.6]   &e[12]\\
\hline
1	&38.42181	&60.39177	& -	&15.79 $\pm$ 0.17	&14.43 $\pm$ 0.08	&13.14 $\pm$ 0.05	&12.40 $\pm$ 0.04	&6.94 $\pm$ 0.05	&13.23 $\pm$ 0.03	&12.25 $\pm$ 0.03	&9.19 $\pm$ 0.05  \\ 
2	&38.83216	&60.43402	&16.00 $\pm$ 0.09	&14.86 $\pm$ 0.08	&13.94 $\pm$ 0.06	&12.28 $\pm$ 0.04	&11.65 $\pm$ 0.03	&7.37 $\pm$ 0.02	&12.45 $\pm$ 0.03	&11.68 $\pm$ 0.02	&8.83 $\pm$ 0.04   \\
3	&39.25507	&60.51685	& -	&15.59 $\pm$ 0.15	&12.96 $\pm$ 0.03	&10.20 $\pm$ 0.07	&9.05 $\pm$ 0.04	&4.00 $\pm$ 0.01	&10.88 $\pm$ 0.03	&8.95 $\pm$ 0.02	&6.58 $\pm$ 0.04 \\
.  &.....&.....&.....&.....&.....&.....&.....&.....&.....&.....&.....\\
\hline
\end{tabular}
\end{table*}

\section{Results \& Discussion}\label{sec:result}
As W4 is located at a low Galactic latitude (l = 134$^\circ$.7, b = 0$^\circ$.9), the contamination due to line of sight stars, 
extra-galactic sources and reddened evolved stars can still be significant at shorter wavelengths although we tried to
 minimize it using various color or magnitude 
cut-offs. Therefore, we adopted different approaches 
to study the clustered and distributed young stellar population in the complex. Since cluster appears as an enhanced 
concentration of point sources in the stellar density map, we used YSO surface density
map of the complex to identify young clusters Contaminating sources in a SFR can be 
eliminated either using spectroscopic observations or using 
deep wide-field optical and X-ray observations as one can evaluate the properties of candidate 
YSOs by comparing their locations in the HR diagram with the theoretical pre-main sequence (PMS) 
isochrones \citep[for details see][]{pan17,pan18}. In the absence of these observations, we primarily use bright MIPS 24 $\micron$ flux 
to identify distributed isolated Class~I population of the complex. This is done mainly to reduce the contamination
 from the background reddened sources and from the faint
evolved sources in our likely sample of distributed mode population. As YSOs emit relatively strong 24 $\micron$ emission during their early evolution phases and also extinction 
effects are less at 24 $\micron$, the MIPS data are particularly important for classification and separation of sources
from the reddened field stars. 
We emphasize that with the aforementioned
photometric approach it is impossible to eliminate line-of-sight PMS sources (if any) those are not part of W4 complex.

\begin{figure*}
\begin{minipage}{0.7\textwidth}
\includegraphics[scale =0.6, trim =0 10 0 20, clip]{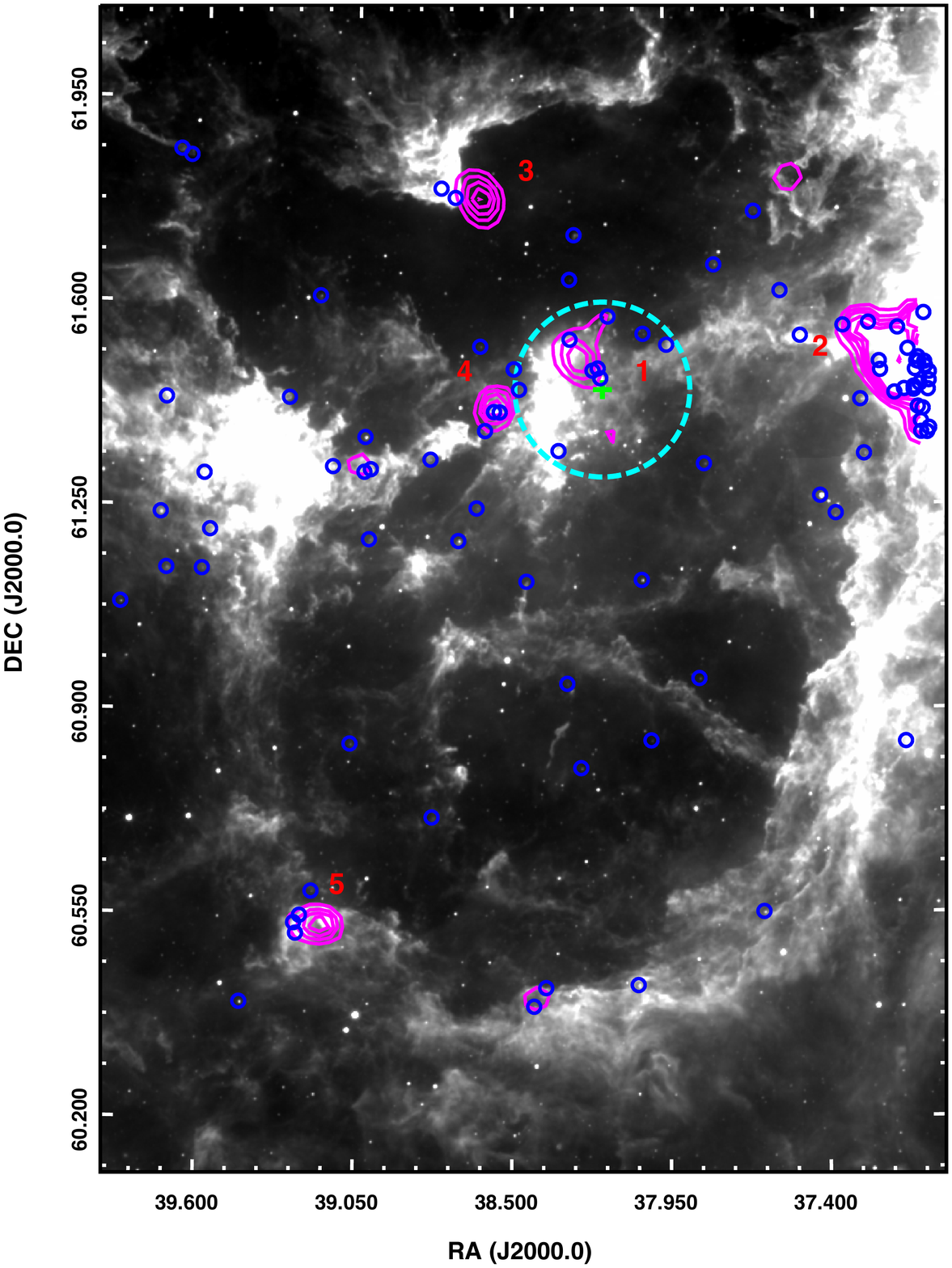}
\end{minipage}
\begin{minipage}{0.3\textwidth}
\includegraphics[scale = 1.0, trim = 200 340 150 350, clip]{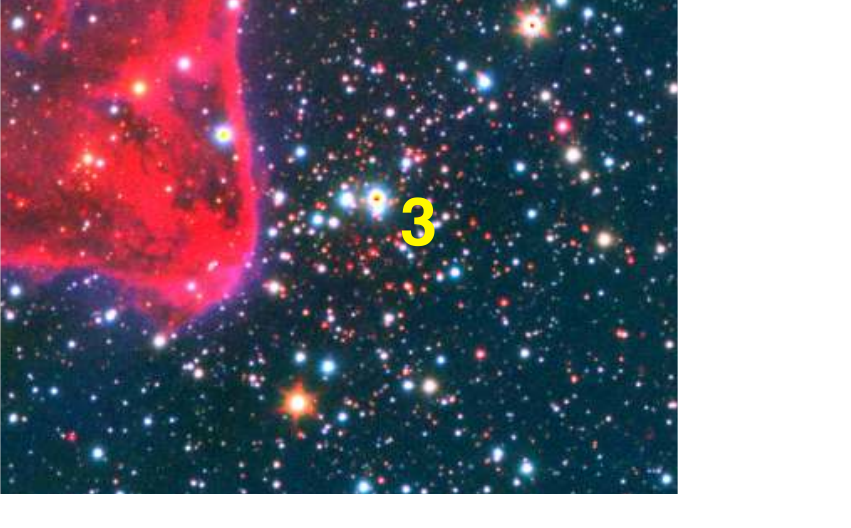}
\includegraphics[scale = 0.78, trim = 180 300 150 320, clip]{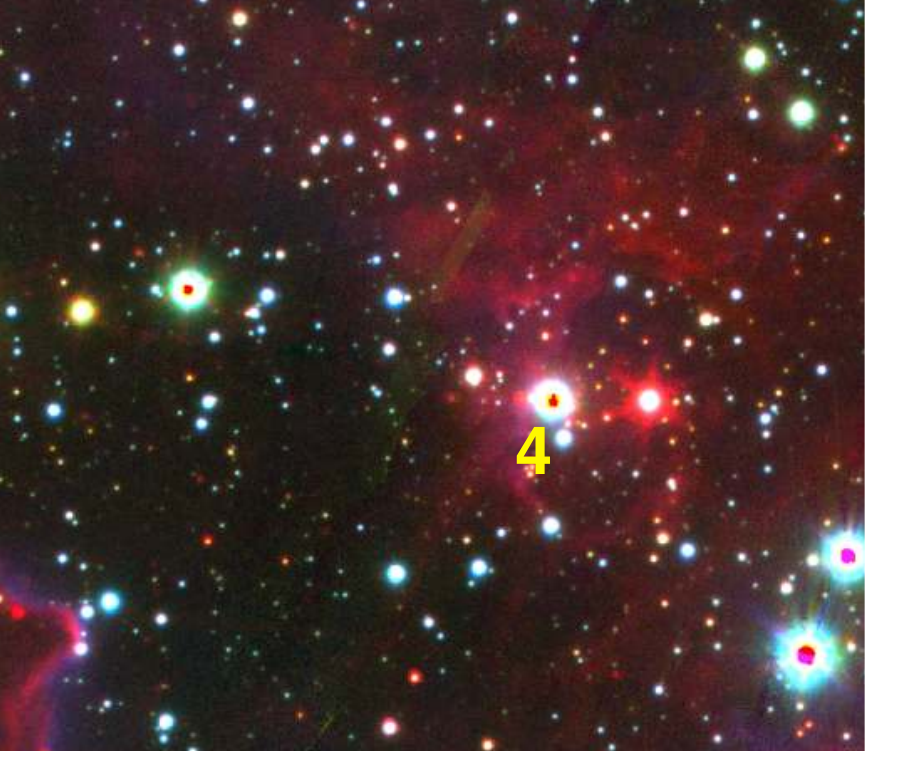}
\includegraphics[scale = 0.66, trim = 150 320 150 270, clip]{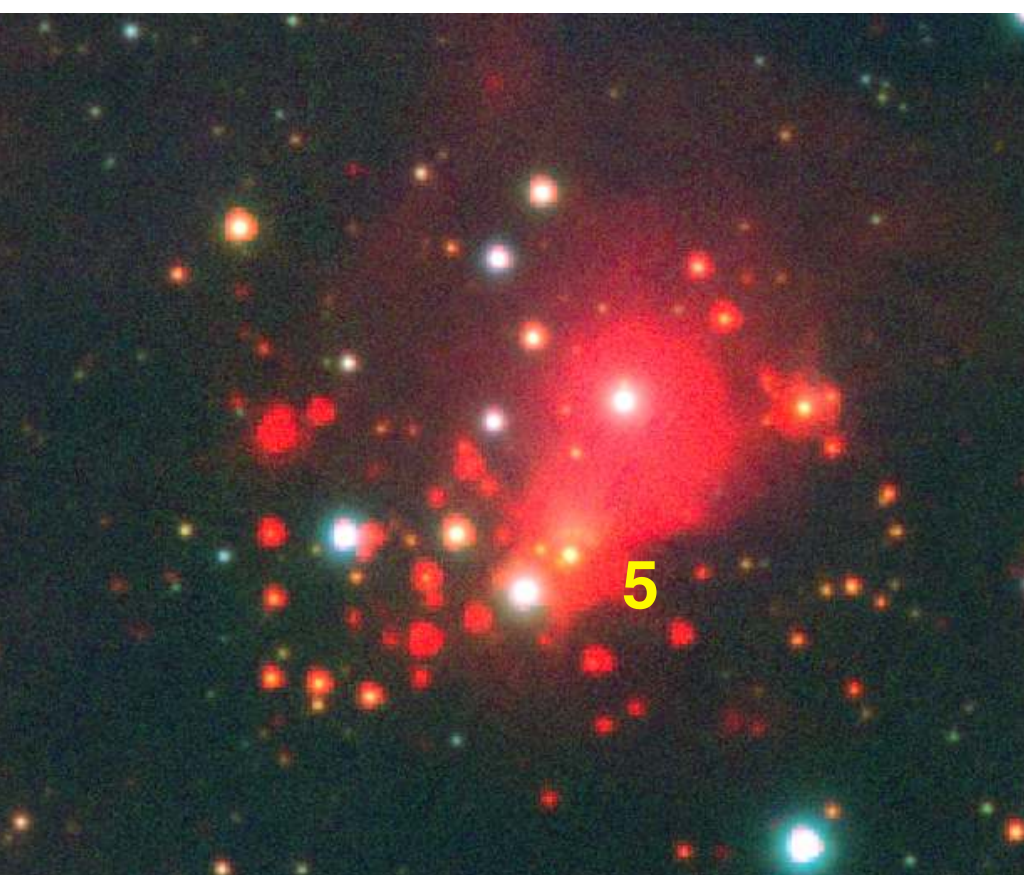}
\end{minipage}
\caption{Left panel: Surface density contours of the identified YSOs overlaid on 22 $\micron$ grey scale image of the W4 H{\sc ii} region.  The contour starts at the level of 
4$\sigma$ with an increment of 1 $\sigma$. 
The blue circles represent the locations of distributed YSOs identified using 24 $\micron$ flux (for details see text). Plus symbol and dashed circle represent the center and the extent of the cluster IC~1805, respectively \citep{pan17}. 
Right panel : zoomed-in view (red: IRAC 4.5 $\mu$m, green : PANSTARR (PS1) y-band and blue : PS1 r-band) of density 
peaks `3'(right-upper), `4' (right-middle) and `5' (right-lower).} 
\end{figure*}
\subsection{Creation of YSO Surface Density Map : Incidence of Clusterings}
We used the nearest neighbor method to generate the YSO surface density map \citep{gut05}. In brief, we divided the region into uniform grids of 1$^\prime$. We measured the radial distance r$_n$(i, j) of the n$^{th}$ nearest neighbour from each position in a uniform grid. 
From this radial distance a surface density is directly computed as $\sigma$$_n$ (i, j)=(n-1)/($\pi$r$_n$$^2$(i, j) 
\citep{cas85}. We varied the value of n and after a series of
experiments decided the value of n = 6, which is a good compromise between the resolution
and the signal-to-noise ratio of the surface density map. 
The contours for the surface density calculated by the above method, plotted above 4 $\sigma$ level, are shown in Fig. 3. 
The surface density distribution reveals five prominent density peaks in the region. One of the density peaks appears at the center of the H{\sc ii} region, i.e., on the cluster IC~1805 (marked as `1'). The second peak is near BRC~5/ AFGL~333, which is studied by \citet{jose16}. The positions of other three peaks do not match with the clusters mentioned in the literature and thus are 
likely new cluster candidates. Below we describe details of these new clusters.  

\subsubsection{Density Peak near to BRC~7}
One of the density peaks (marked as `3' in Fig. 3) is at $\alpha$$_{2000}$=38$^\circ$.621, $\delta$$_{2000}$=+61$^\circ$.787. It is near the apex of BRC~7. 
To study the stellar extent of this clustering, we 
plotted its radial density profile (RDP) using our YSO sample. We divided the cluster into a number of concentric
circular regions. To calculate the projected stellar density in each concentric annulus, we divided the
number of stars within each annular region by the respective area. 
The densities thus obtained are plotted as a function of radius in Fig. 4a. 
The  error bars are derived assuming that the number of stars in each annulus follows the Poisson
statistics.

We model the RDP with a function similar to the Elson, Fall \& Freeman (EFF) profile \citep{eff87}, often used to characterize the radial profiles 
of young clusters in the galaxies. The fitted function (dashed curve in Fig. 4a) is of the form: 
\begin{center}
$\rho (r) \propto {\displaystyle{\rho_0} [1+(\frac{r}{a})^{\beta}]^{-{\gamma}/\beta}}$,
\end{center}

where $\beta=2.0$ \citep{eff87}. $\rho_0$, a and $\gamma$ are the central surface density, 
 measure of the core radius and the power-law slope at large radii, respectively. 
Similarly, we also fit the observed RDP with the empirical King profile \citep{kin62} of the form:  
\begin{center}
$\rho (r) \propto {\displaystyle {\rho_0} \{[1+(\frac{r}{r_c})^{2}]^{-0.5} - [1+(\frac{r_t}{r_c})^{2}]^{-0.5}\}^{2.0}}$,
\end{center}
 where the parameters $\rho_0$,  $r_{c}$ 
and $r_{t}$ correspond to the central density, core radius and tidal truncation radius, respectively. The core radius (r$_c$) 
provides information about the innermost cluster structure whereas r$_t$ and $\gamma$ tell about the outermost region of the cluster.
 The empirical King profile fitting on the observed RDP is shown with solid black curve in Fig. 4a. In present work, we are only interested in the outer extent of the clusters. The radius where the stellar density profile merges to the background density is defined as the
 cluster radius ($r_{cl}$).  
 Based on the RDP, we find that $r_{cl}$ (where the surface density becomes nearly constant) 
is about 3$^\prime$.

A massive star LSI +61 298 \citep[B1 V;][]{reed05} lies within the cluster radius. The radial velocity 
of the star is -32.9 km/s \citep{hua06}, consistent with the velocity of the molecular cloud (-55 to -32 km/s) associated with the W3/W4 complex \citep{hey98,car00}, thus indicating that the cluster is associated with the region. 

\begin{figure*}
\includegraphics[scale = 0.5, trim = 0 30 30 40, clip]{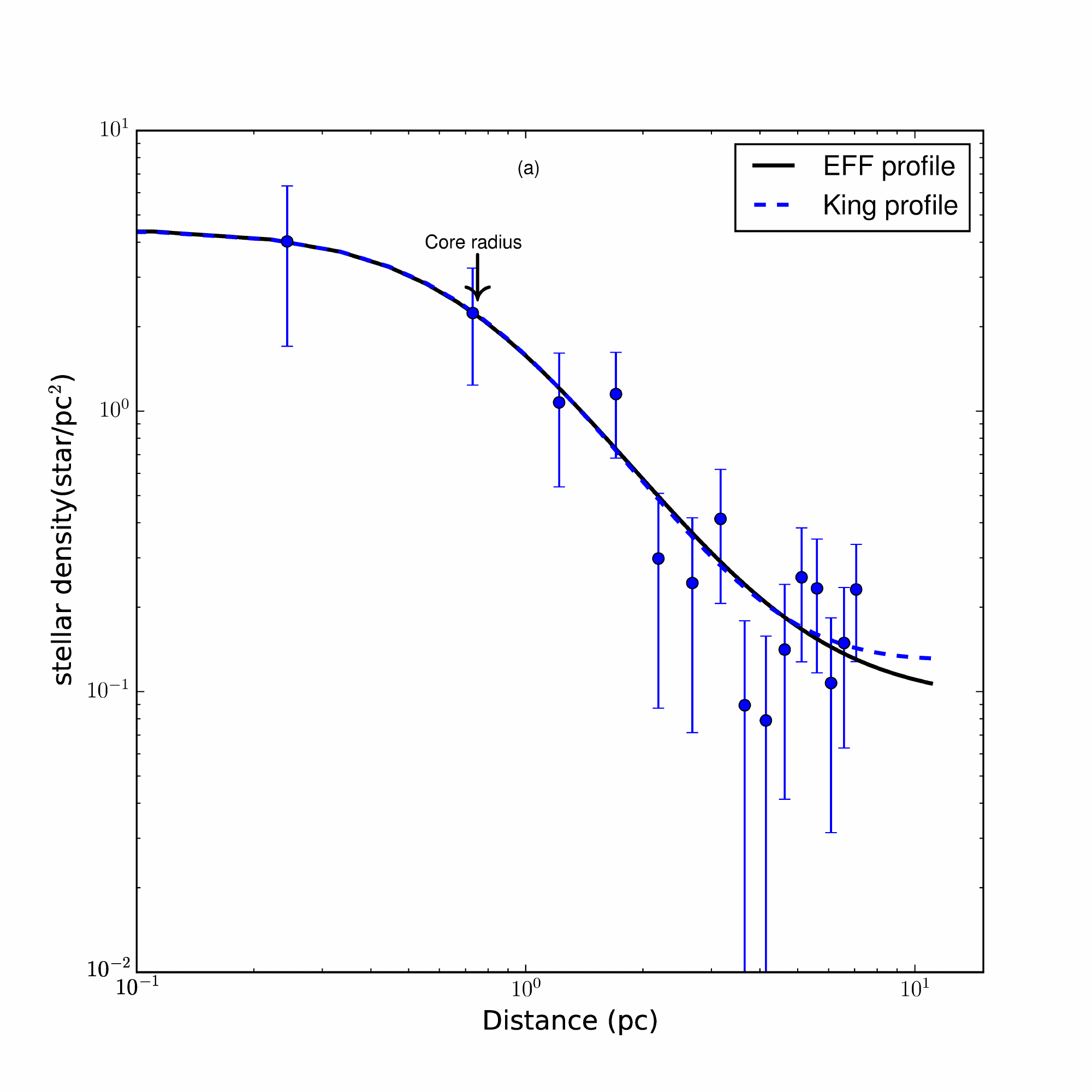}
\includegraphics[scale = 0.5, trim = 0 30 30 40, clip]{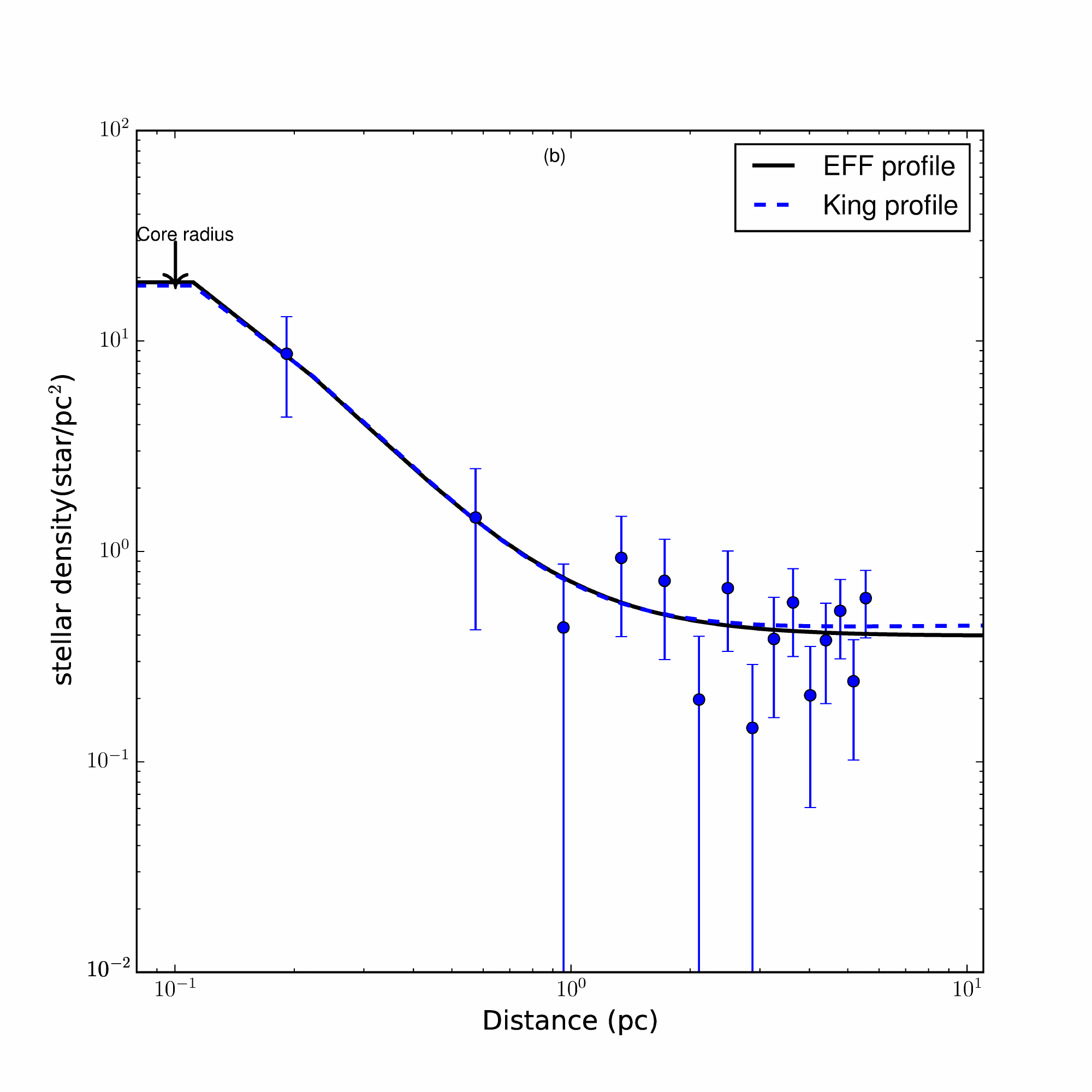}
\includegraphics[scale = 0.5, trim = 0 30 30 40, clip]{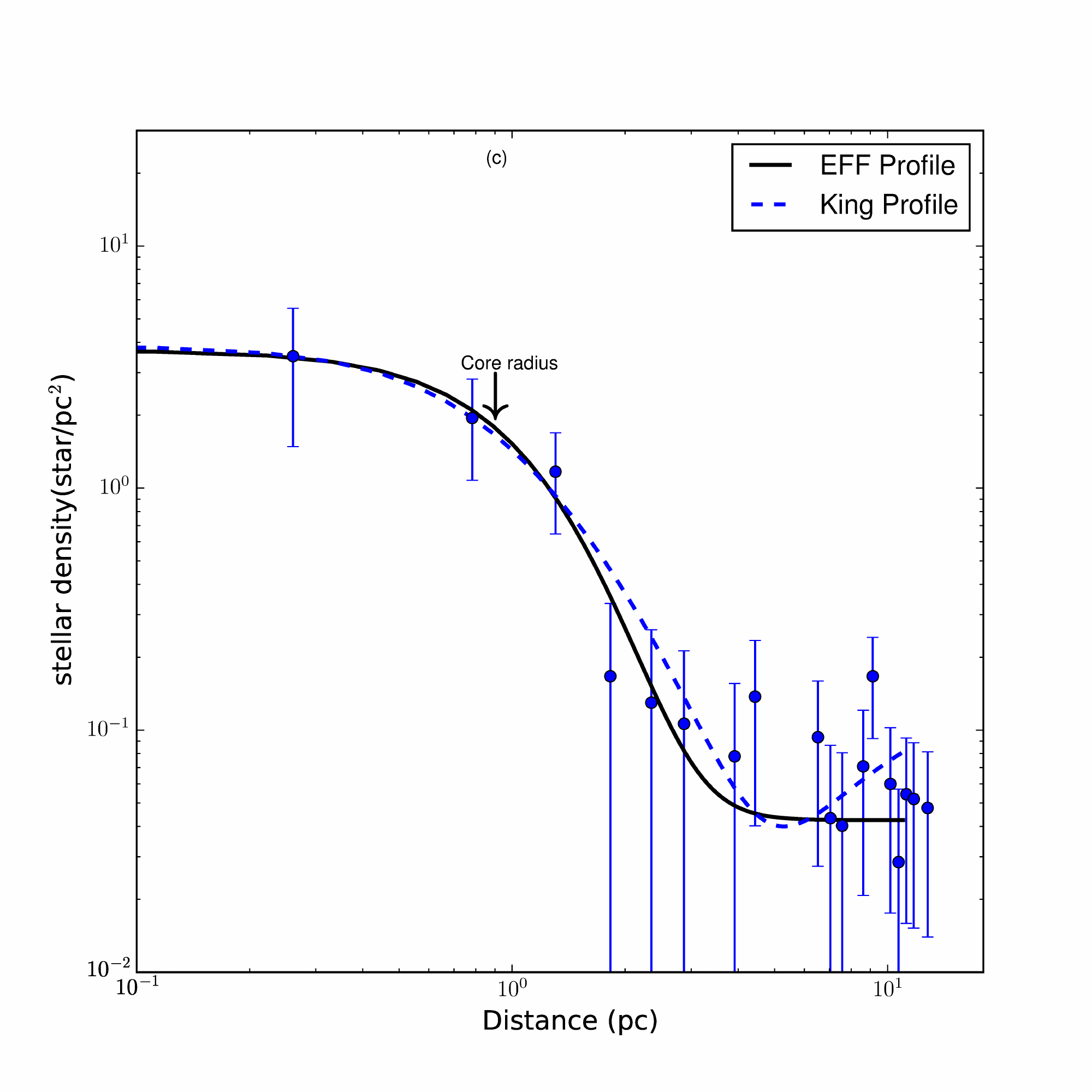}
\caption{Stellar surface density as a function of radius for the YSOs within the density peaks (black dots) marked as `3' (a), `4' (b), `5' (c) in Fig. 3. Also, the best-fit King profile in solid black and the best-fit EFF profile in dashed blue for the radial profiles 
of the density peaks are plotted.  }
\end{figure*}

\subsubsection{Density Peak between CG~7S and IC~1805} 
One of the density peaks is near the cometary globule CG~7S ($\alpha$$_{2000}$=38$^\circ$.567, 
$\delta$$_{2000}$=+61$^\circ$.421) which is marked as `4' in Fig. 3. We also plotted the 
RDP for this clustering using our YSO sample and fitted with the EFF and King  profiles (Fig. 4b). The radial extent 
of the cluster is $\sim$ 2 arcmin. This clustering also contains a massive star LSI~+61~297 which is 
designated as a B1 V type in $SIMBAD$. GAIA has measured a parallax of 0.411 $\pm$ 0.035 mas for this 
star, placing it at a similar distance as reported for the H{\sc ii} region \citep{gaia18}. 
The cometary globule CG~7S was extensively studied by \citet{lef95} in order to test the RDI model. 
Using $^{12}$CO, $^{13}$CO and C$^{18}$O  observations they found a small 
dense core hidden by a low density halo. They compared their observations
with the numerical simulations of the RDI model and suggested that the globule is in the 
state of re-expansion, exhibiting the features of the pre-cometary phase. 
The suggested typical age of the globule was $\sim$ 1.4 Myr which is less compared 
to the age ($\sim$ 2.5 Myr) of the cluster IC~1805.

\subsubsection{Southern density Peak} 
Towards the south-east periphery of the H{\sc ii} region, we notice another YSO density peak (marked as `5'; $\alpha$$_{2000}$=39$^\circ$.179, $\delta$$_{2000}$=+60$^\circ$.537). 
The RDP for this clustering yields a cluster radius of $\sim$2.5 arcmin (see Fig. 4c). This clustering
 does not contain previously known massive star but there is a massive OB type star 
$\sim$ 6 arcmin away from the cluster center \citep{reed05}, probably not the 
cluster member. The integrated CO (1-0) emission map by \citet{hey98} also shows a strong CO emission 
towards the south-east side of the H{\sc ii} region which matches with the location of this high density 
YSOs peak. The V$_{LSR}$ of this small molecular cloud is $\sim$ -43 km/s,
consistent with the molecular cloud velocity (-55 to -32 km/s) associated with the W3/W4 complex \citep{hey98}. 
The strong CO emission and apparent clumping of YSOs suggests that 
this cluster is associated with the H{\sc ii} region. 
\subsubsection{Evolutionary stages of the newly identified clusters}
\begin{figure*}
\includegraphics[scale = 0.8, trim = 0 150 0 250, clip]{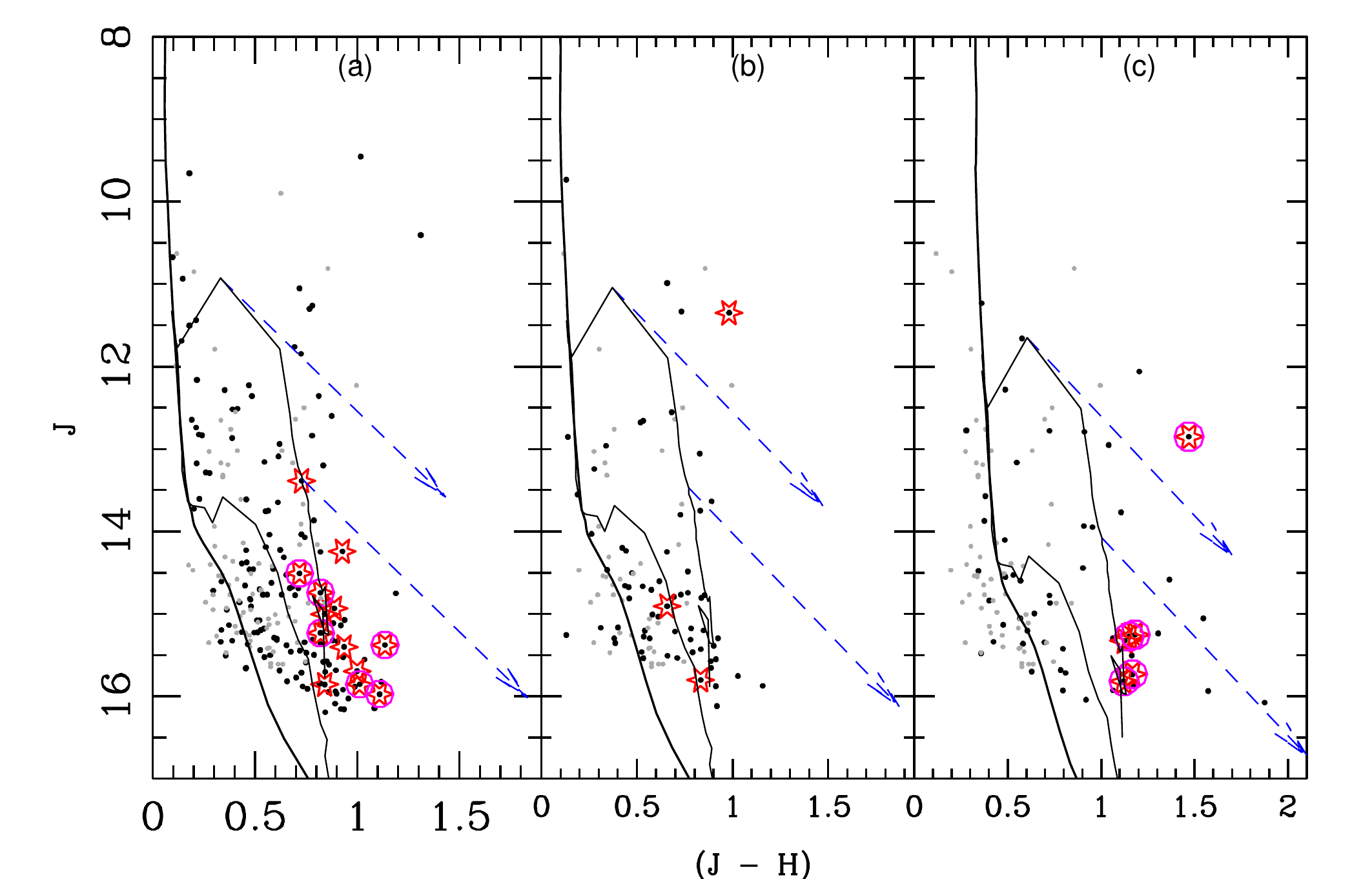}
\caption{ 
$J$ vs. ($J$-$H$) color-magnitude diagrams for the stars within the density peaks (black dots) marked as `3' (a), `4' (b), `5' (c) in Fig. 3. PMS isochrones from \citet{sie00} for 1, 10 Myr and MS from \citet{gir02}, corrected for the adopted distance and reddening, are also overplotted. Open circles and red stars represent candidate YSOs within the core radius and cluster radius, respectively. The upper and lower 
dashed lines are reddening vectors of A$_V$ = 10 mag for 1 Myr PMS stars of 4 M$_\odot$ and 2 M$_\odot$, respectively. }
\end{figure*}

To study the evolutionary stages of the new clusters, we used the 2MASS NIR $J$ vs. ($J$ - $H$) CMDs. The CMDs for the stars in the newly identified clusters are shown in Fig. 5.  We note that the YSOs shown in the CMD are biased to the low-extinction region only. Thick black dots are the sources within the radial extents of new clusters. To compare the contribution 
of field stars in YSO population of the new clusters, we took a field region towards the northern side of the H{\sc ii} region. 
 The small grey dots in Fig. 5 represent 
the sources in the field regions of the same areas as the cluster regions.  
 The star symbols and open circles represent the YSO candidates within the cluster radius and core radius, respectively. In Fig. 5, PMS isochrones 
of 1 and 10 Myr from \citet{sie00} and main sequence (MS) from \citet{gir02}, corrected for the adopted distance (2.0 kpc) and average reddening (obtained from the extinction map) are also plotted. 
The upper and lower dashed lines represent reddening vectors of $A_V$ =$10$ mag for 4 M$_\odot$ and 
 2 M$_\odot$ PMS star, respectively. 
The CMDs show that most of the 
identified YSOs in the newly identified clusters are $\le$ 1 Myr old. 
Due to low statistics of the sources, it is impossible to obtain statistically robust  ages of the clusters using $J$/($J$ - $H$) CMD. However, as can be seen
from the CMDs, a majority of the YSOs are located beyond 1 Myr
PMS isochrone, implying that the ages of the clusters likely to be less than 
1 Myr.   Since most of the YSOs in these clusters are younger compared to the central cluster, i.e., IC~1805, 
these clusters belong to a different generation of star-formation in the complex. 
We discuss the origin of such clusters in Sect. 4.2.3. 

\subsection{Distributed Young Stellar Population and its Origin}
\begin{figure}
\includegraphics[scale = 0.9, trim = 80 50 0 480, clip]{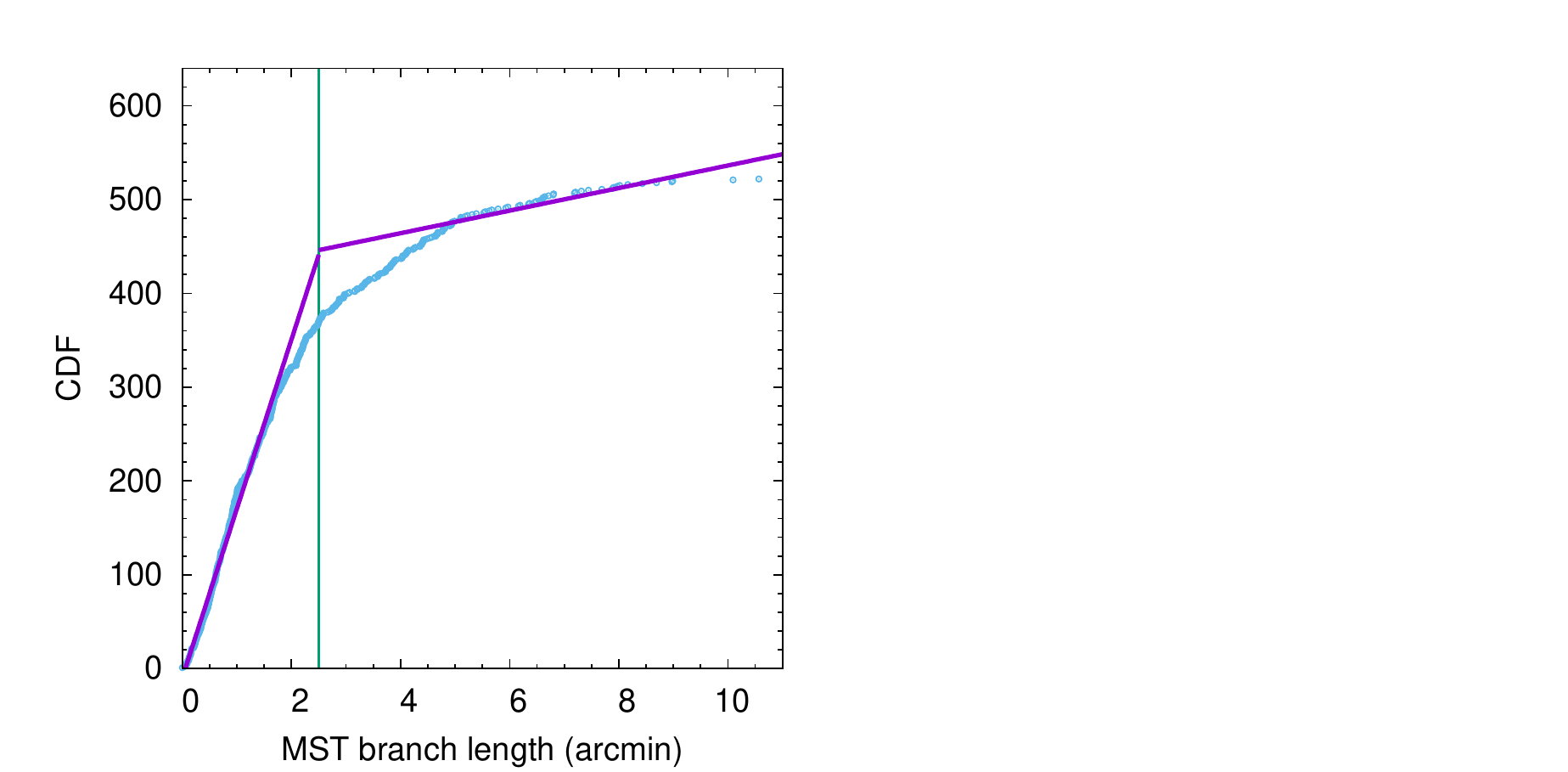}
\includegraphics[scale = 0.9, trim = 80  50 0 480, clip]{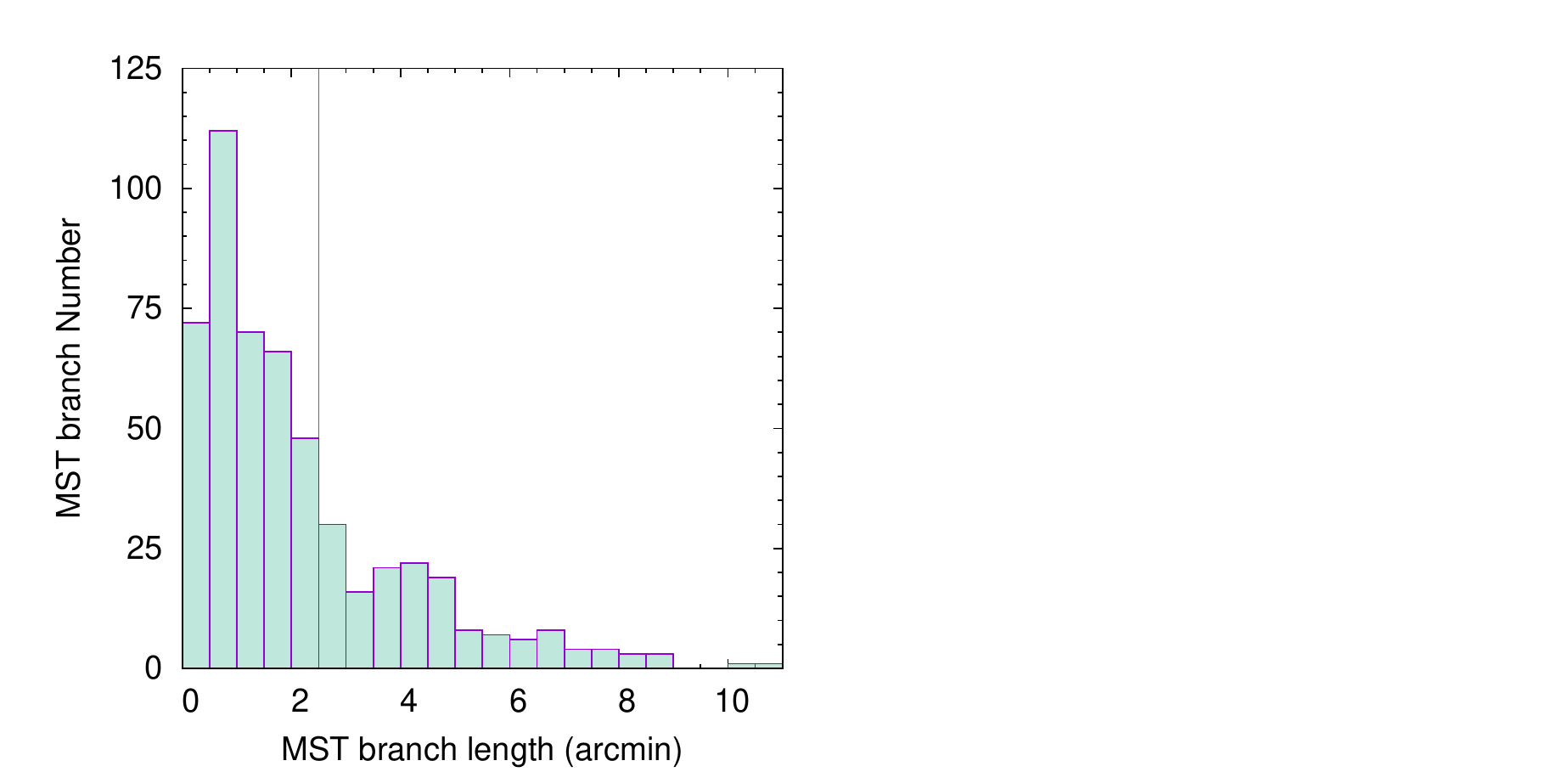}
\caption{CDF and histogram of MST branch lengths used for critical length analyses of the 
YSOs. The CDF plot has sorted length values on the X-axis and a rising integer counting index on 
the Y-axis. The solid line is a two-line fit to the CDF distribution. The vertical line 
stands for the critical lengths obtained for the core region.}
\end{figure}

Figure 3 shows the W4 complex at 22 $\micron$ along with the YSO candidates
identified using 24 $\micron$ MIPS data. The center of the cluster IC 1805 and its extent
are also marked on the image as a plus symbol and the dashed circle, 
respectively. As can be seen, most of the MIPS identified YSOs show scattered distribution 
except an enhanced concentration (or group) at the western border of the bubble. 
Most of these sources are  primarily located beyond the cluster radius. We note that in a recent work 
\citet{sung17} suggested that the cluster radius is likely to be $\sim$ 15 arcmin ($\sim$ 10 pc). 
Even if we assume a cluster radius of 10 pc, we find $\sim$ 90\% of the
24 $\micron$ based class~I YSOs located beyond this radius. 
 Among these, around 50\% are likely scattered as they are not spatially coinciding with any clusters
found in the surface density map. 
We note that the group of 
YSOs found at western border of the bubble are part of a filamentary cloud \citep[for details, see][]{jose16}. 
We also obtained characterization of the scattered distribution of the YSOs from the minimum spanning tree (MST) method as explained below. 

\begin{figure}
\center
\includegraphics[scale = 1.1, angle=90, trim = 50 150 480 30, clip]{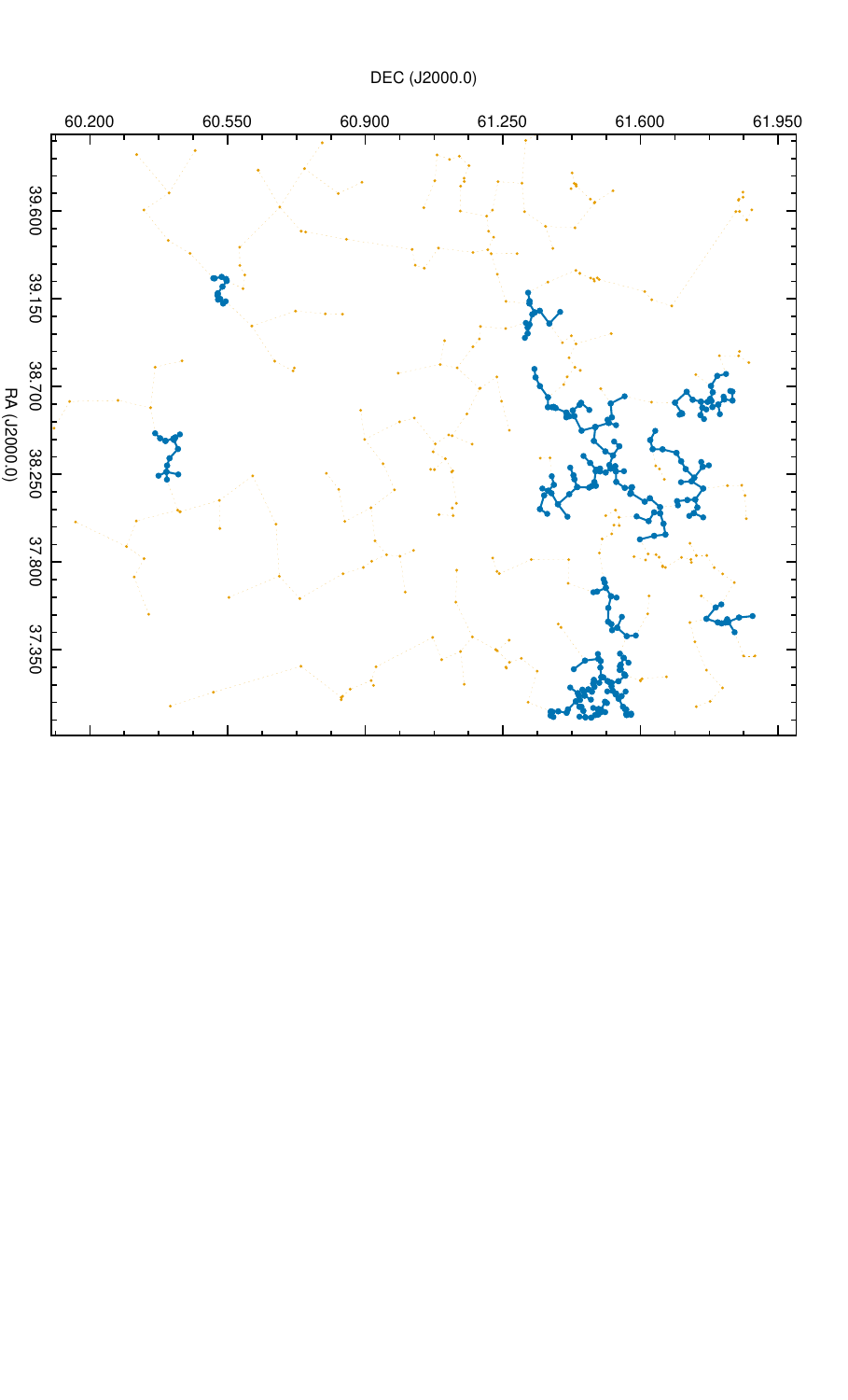}
\caption{MST for the MIPS identified YSOs in the W4 region. The black dots
connected with solid black lines are the branches smaller than the critical length for
the cores.}
\end{figure}

A scattered and clustered distribution of the
sources can be obtained from the MST of
the source positions \citep{sch06,gut09,saur16}. The MST is defined as the network of lines,
or branches, that connect a set of points together such that the
total length of the branches is minimized and there are no closed
loops. To obtain the MSTs for the YSOs in the region, we used the approach
as suggested by \citet{gut09}.
The histogram of the MST branch length, shown in Fig. 6 (right panel), reveals a peak at short spacing and a long
tail towards large spacing. In Fig.
6 (left panel), we plot the cumulative distribution for MST branch
lengths.  The cumulative distribution function (CDF) shows a three-segment curve; a steep-sloped
segment at short spacings, a transition segment that approximates
the curved character of the intermediate length
spacings, and a shallow-sloped segment at long spacings \citep[cf.][]{gut09}.
We define a core region
where majority of the sources are found
in small spacing, i.e., steep segment. We fitted two straight lines to
the shallow and steep segments of the CDF and adopted their intersection as
the MST critical branch length (see Fig. 6).
The core and the extended  
regions were isolated by clipping the MST branches longer
than the critical length described above.
The black dots connected with black lines
are the branches smaller than the critical
length in the core region. The scattered YSO population (the YSOs outside the cores)
clearly show up in the region (see Fig. 7). Since both the above approaches use different methods to identify clusters, some discrepancy on clustering properties 
is likely. However, all the clusters seen in the surface density map also appear in the MST analysis. Though some new small groups appear in the MST plot, 
they appear weak in the surface density map. We considered only those groups as cluster that have strong density 
enhancement in the surface density maps as well as MST plot. We found that the fraction of the distributed YSO
 population (the YSOs outside the cores) 
is $\sim$ 50\% of the total YSOs in the region. 
If we assume that most of these young sources are part of the W4 complex, it
is intriguing how these sources might have formed. Therefore, we evaluate a few processes 
that can lead to distributed mode of star formation in a star forming complex. 

\subsubsection{Hierarchical Fragmentation and Evolution of Molecular Cloud}
In view of the recent numerical simulations of cloud formation and evolution, young star-forming regions are formed by global  hierarchical
gravitational collapse \citep{sem09,avi12,sem17}. In this scenario, as 
molecular clouds are often turbulent \citep{kle98,kle00a} the 
hierarchical structures within the dynamically evolving cloud are created. The free-fall time of any self-gravitating perturbation in the structure is shorter 
than the overall dynamical time of the system. 
As a result, collapse of molecular cloud is hierarchical, i.e., consisting  of small-scale collapses within larger-scale ones. 
The larger-scale collapse culminate a few Myr later than the small-scale collapse. In other words, star formation occurs in small-scale clump/cores 
before the onset of star formation in the large-scale clump preferentially located at the centre of the cloud potential. 
As shown in \citet{sem17}, massive star or cluster is preferentially located at the center of 
the cloud while the low-mass old stars are located at the outskirts.  
This process may results in older stars confined to a region where new protostars are forming.  
Recently, \citet{pan18} also 
observed a similar age segregation in young cluster Berkeley 59. As shown in the simulations, 
matter mainly channels to the center of potential through hierarchical network of filaments, thus  the older stars formed in the  
dense structures of the filaments 
share the in-fall motion of the gas on to the central clump \citep{sem17}. They have larger velocities and tend to be distributed over 
larger areas than the younger stars formed in the central clump. If we assume that the aforementioned scenario
is responsible for the scattered distribution of the YSOs at the outskirts of the complex, one would expect these 
sources to be older than the central cluster. Given the fact that many YSOs are Class~I/II stars of age likely 10$^{6}$ yr \citep[e.g., see discussion in][]{eva09}, this process, therefore, should 
not be a major cause for the origin of such sources. 
\subsubsection{Dynamical Ejection of Stars From Cluster Forming Clumps}
Several mechanisms have been proposed for the formation of 
low-mass stars and brown dwarfs. One of them is the 
dynamical ejection \citep{rei01,boss01,bat02,good07} which suggests that low-mass objects (e.g., brown dwarfs) 
form in dense multiple systems, and they are ejected due to dynamical instabilities 
before they can accrete more mass from their environment. These sources would have a higher
velocity dispersion than higher mass objects.  

The detections of outflows and disks around brown dwarfs suggest that brown dwarfs and low-mass stars 
share a similar formation process. If we assume that a fraction of the low-mass stars 
in a clustered environment followed the ejection scenario proposed by 
\citet{rei01} with a resulting average velocity larger than
the cluster escape velocity or with a velocity dispersion larger than a few km/s as found in simulations, 
then a large number of low-mass stars are expected to be lost relatively quickly. 
For example, 
\citet{kro03} suggested that the ejection mechanism could explain a part of the deficit of 
substellar objects in the Taurus association.  
Extreme dynamical interactions, namely, close encounters
between stars in multiple systems, can produce high-velocity
runaway stars. Ejection of massive stars with a high velocity from
cluster center has been noted in many cases, e.g., 
\citet{ster95} have calculated time-scales 
for close systems  and predicted velocities of 3-4 km/s for 
the ejected stars. Assuming an ejection velocity of $\sim$ 3 km/s  perpendicular
to the line of sight, a distance of 8.5 pc could be 
traversed in 3 Myr which is the upper-age limit of the cluster IC 1805. 
However, in the present work, we find most of the 24 $\micron$ YSOs are distributed $>$ 17 pc away from the cluster center.  
Thus it is natural to hypothesize that the low-mass young stars observed at the outskirts of the complex 
are unlikely the ejected members of the cluster IC 1805. However, if we assume that the low-mass stars are the
ejected members of the small-scale cores/clumps that formed during the initial stage of
global hierarchical gravitational collapse of the cloud, one would expect these sources to be older
compared to stars at the cluster center. However, we acknowledge that without precise knowledge of age and motion of the sources, a conclusive answer can not be given.
\subsubsection{Effect of Stellar Feedbacks: star formation in small-scale structures such as pillars and globules}
\begin{figure*}
\includegraphics[scale = 0.82, trim = 0 100 0 100, clip]{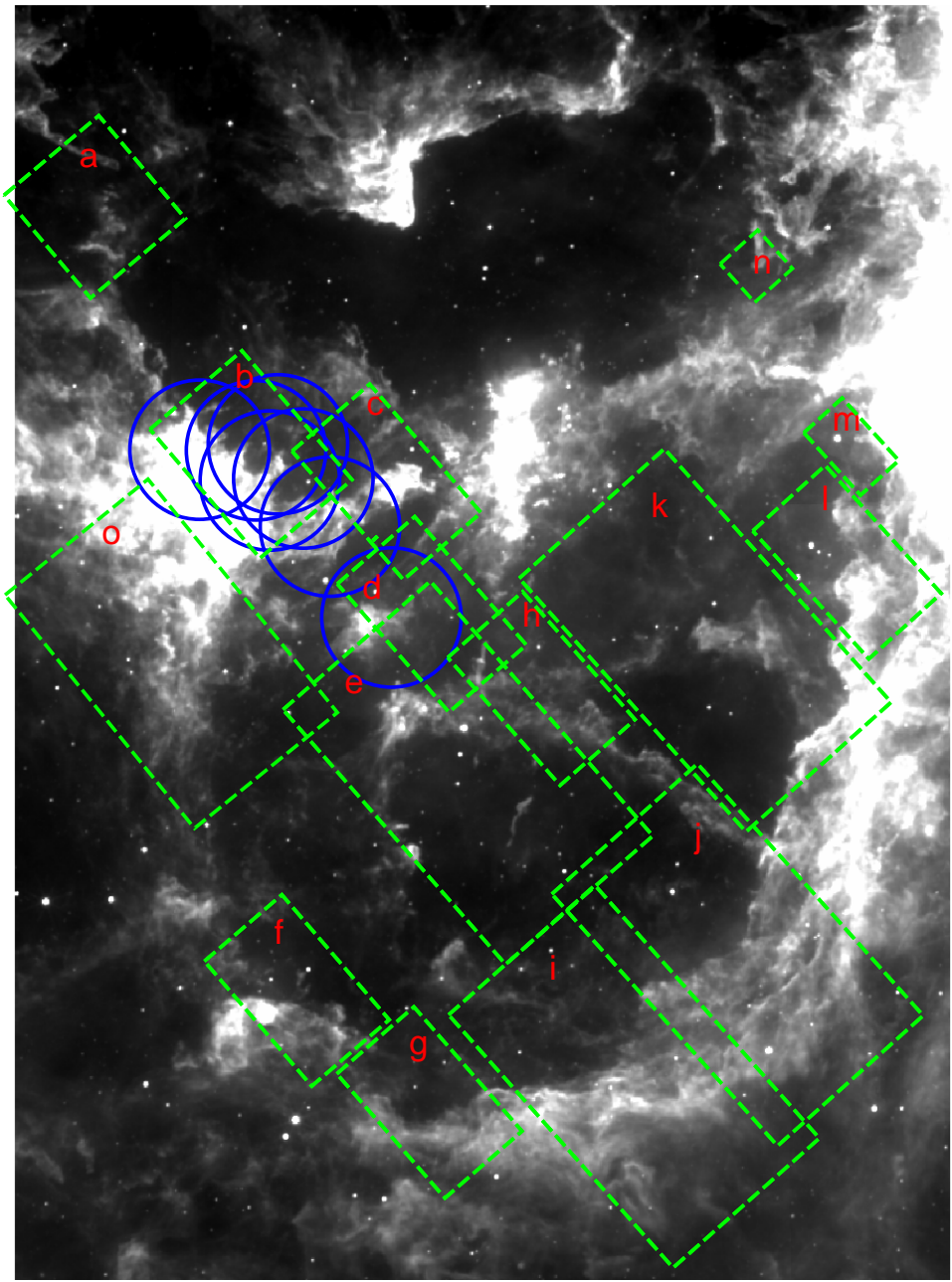}
\caption{The WISE 12 $\micron$ image of the H{\sc ii} region. Rectangular regions enclose the ETLSs with stars at their tips. The area covered by 
\citet{gah07} to identify globulettes are shown with circles.}
\end{figure*}
\begin{figure*}
\includegraphics[scale = 0.9, trim = 10 206 0 180, clip]{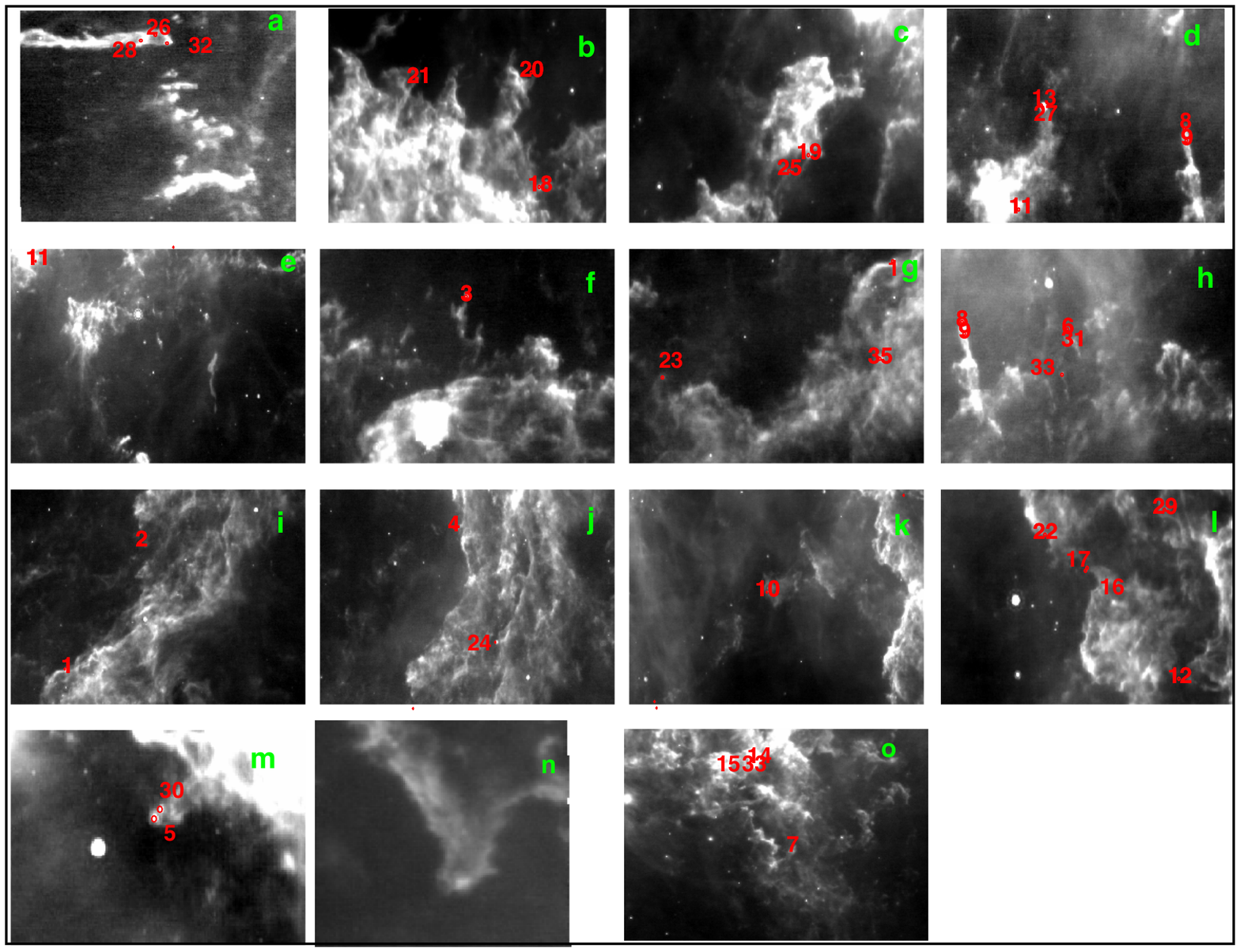}
\includegraphics[scale = 0.3, trim = 0 0 0 0, clip]{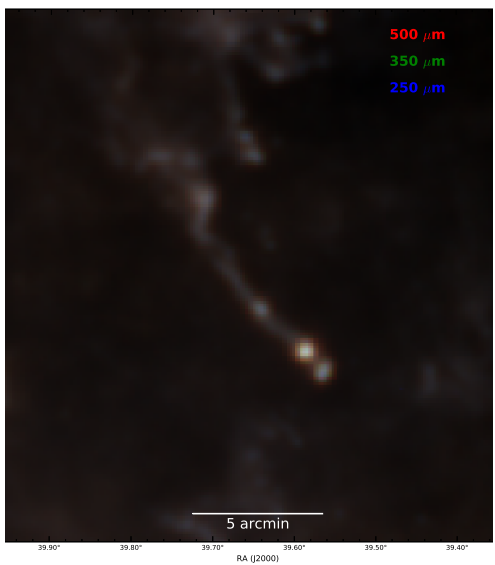}
\caption{Zoomed-in view of the ETLSs within the rectangular regions of Fig. 8 (upper panel) and an ETLS prominent at longer wavelength with two Class~I sources  
at its tip (lower panel).}
\end{figure*}

Massive stars play a critical role in the evolution of natal molecular cloud by regulating the local ISM through their intense radiation and winds \citep{lop11,dale13,wal13,jos13,sam14}. 
Recent simulations  and observations \citep[e.g.,][]{mye09,schn13,mali13,kirk14,sam15,rayn17,dutt18}
 show that cluster formation mainly occurs at the center of cloud potential
to which gas flows through a network of threaded filamentary or low-density striation like structures. 
 The erosion of these filaments/accretion flows due to the 
expanding radiation and winds from the massive stars of the cluster results in elephant-trunk or pillars \citep{dale15}. 
Elephant trunks are long columns of gas and dust that extend radially from the
molecular shell driven by the massive stars toward the central star or cluster 
and have thin bright rim-like structure at their heads.
Pillars have typical widths of 0.1 - 0.7 pc and size scales of $\sim$ 0.6 - 3 pc, generally attached to their natal 
molecular cloud whereas globules are isolated and have a head-tail appearance \citep{gah06,schu06,sch16}. 
Generation of such 
structures also depends on the fractal dimension of the cloud. 
For example, using SPH simulations, \citet{wal13} have shown that 
if the fractal dimension of the molecular cloud is low, the border of the H{\sc ii} region 
is dominated by extended shell-like structures, which break up into a small number 
of massive high-density clumps to spawn star clusters. Conversely, if the fractal dimension 
is high, the border of the H{\sc ii} region is dominated by a large number of pillars and 
cometary globules, which contain compact dense clumps and tend to spawn single stars or 
individual multiple systems;  the stellar masses are somewhat lower, and the stars are more
widely distributed. 

The number of such structures in a given complex depends on the hierarchical nature of star-forming cloud. 
We note that elephant trunk-like 
structures (ETLSs) are explained based on the models that do not require clumpy material initially \citep[e.g., hydrodynamical instabilities
  and high curvature in the shell;][]{cha11b,trem12b,trem12a}. 
Since the masses of the clouds that are associated to the ETLSs are often small ($\sim$ 5 - 10 M$_\odot$),
they are likely sites of low-mass star formation. Therefore, finding elephant- and pillar-like structures are 
of great importance for understanding the formation and distribution of low-mass stars 
in massive star forming complexes. 

W4 H{\sc ii} region powered by several O-type  stars of the cluster IC~1805 
is also a reservoir for such objects. The stellar feedback has significantly influenced this complex as it contains many BRCs. Out of these BRCs, BRC~5 and BRC~7 are extensively investigated and the small-scale star formation events 
triggered by the hot massive stars are suggested \citep{ogu02,fuk13,pan14,jose16}. 
We suggest that the stars associated with these BRCs and the newly identified clusters
are a result of the compressing effect of massive stars on the in-falling molecular material to the cluster potential
through relatively massive filaments \citep[see simulations of][]{dale12}. 
The MIR/FIR color-composite image of W4 H{\sc ii} 
region is shown in Fig. 1. It reveals several thin and elongated
ETLSs within the H{\sc ii} region that look morphologically different from usual BRCs. These structures 
have their tips pointing towards  the massive stars, suggesting  that their formation and evolution is strongly   
influenced by the massive members of the cluster IC~1805. Similar structures have been observed in the high-resolution 
images of many H{\sc ii} regions taken with the Hubble Space Telescope and the $Spitzer$ Space Telescope \citep[e.g.][]{hest96,smi10,cha11b}. 

Due to low-resolution of $Herschel$ images, we used $Spitzer$ images to identify ETLSs. Based on the $Spitzer$ images, we have identified several ETLSs in the complex. Out of these $\sim$ 24 ETLSs seems to have stars at their tips.  
The locations of the identified ETLSs and the zoomed-in view of the 
ETLSs alongwith stars at their tips are shown in Fig. 8 \& 9, respectively. A zoomed-in view of one of the ETLSs at 
longer wavelength is shown in Fig 9b, where one can see two Class~I sources at its tip. The positions of the stars at the tips of the ETLSs are catalogued in Table 2. We note that the ETLSs identified by us are elongated structures with aspect ratios
in the range of $\sim$ 2 - 7 and lengths of about 0.4 - 2.0 pc.  In total we have identified $\sim$ 38 ETLSs of lengths greater than 0.4 pc.
However, we acknowledge that our identification is biased towards longer ETLSs. We are looking at the ETLS in 2D projection, thus their sizes and total numbers are likely an underestimation. We also note that the cold
gas content of  many small-scales ETLSs
might have  been completely cleared by the UV photons of the massive stars. Therefore, some of the YSOs seen today without associated pillars could be part
of such a process.    
\begin{figure}
\center
\includegraphics[scale = 0.4, trim = 0 120 0 50, clip]{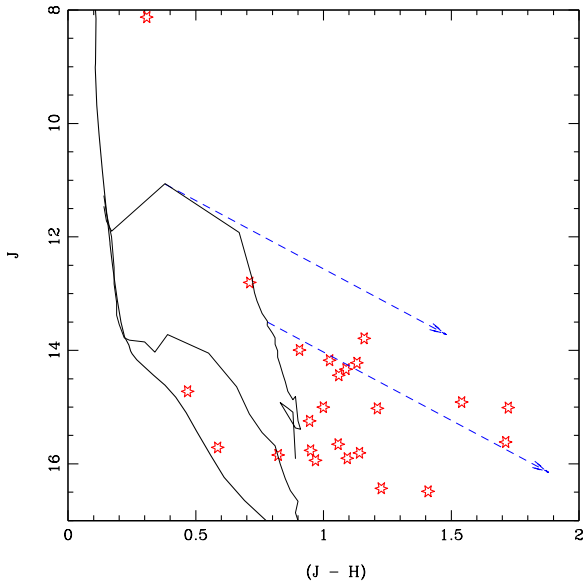}
\caption{The $J$/($J$-$H$) CMD for the stars at the tips of ETLSs. PMS isochrones from \citet{sie00} for 1, 10 Myr and MS from \citet{gir02}, corrected for the adopted distance and reddening, are also overplotted. The upper and lower 
dashed lines are reddening vectors of A$_V$ = 10 mag for 1 Myr PMS stars of 4 M$_\odot$ and 2 M$_\odot$, respectively.  }
\end{figure}
\begin{figure}
\center
\includegraphics[scale = 0.4, trim = 0 120 0 50, clip]{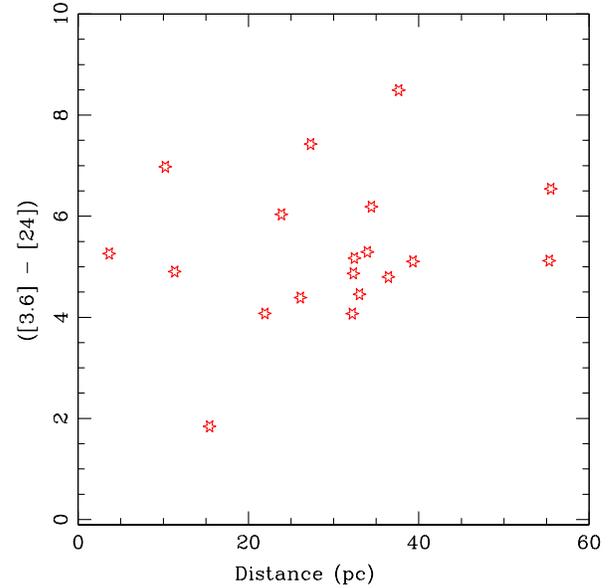}
\caption{The ([3.6] - [24]) colours of the ETLSs stars as a function of radial distance from the 
center of the cluster IC~1805. Most of these sources have [3.6]-[24] color greater than 4.5, thus are likely protostars \citep[e.g., see][]{gut09}.}
\end{figure}

\citet{cha11b} found YSOs at the tips of many ETLSs in W5 E H{\sc ii} region. 
Based on the optical CMDs and SED fitting of these stars, they found  mass $<$ 2 M$_\odot$ for most of the 
stars. 
In order to determine the masses of the YSOs located at the tips of the ETLSs identified in the present work, we used 
$J$ vs. ($J-H$) diagram. This is because among near-infrared bands, the effect 
of the excess emission on the mass estimation due to circumstellar disk in the J-band is minimal \citep[e.g.,][]{mey97,luh03b}. 
Fig. 10 shows $J$ vs. ($J$ - $H$) CMD of the stars at the tips of the ETLSs. For comparison, we have 
also over-plotted the PMS isochrones for 1 Myr and 10 Myr from \citet{sie00} and MS from \citet{gir02}, 
corrected for the adopted distance and average reddening. The location of the stars in the CMD 
clearly indicates that most of them are very young ($<$ 1 Myr old) low-mass objects. 
Fig. 11 shows the  ([3.6] - [24]) color vs. distance (from cluster center) of the stars associated with the ETLSs. 
As can be seen, stars at the tips of ETLSs identified in the present work are lying at a distance range 
of 5 to 90 arcmin from the cluster center, with majority beyond 30 arcmin ($\sim$ 17 pc). Their distribution suggests
that most of these young stars are lying at the outskirts of the complex. If it is to be believed that these 
stars are formed due to compression and squeezing of pre-existing material like star formation in BRCs,
it can be argued that small-scale star formation is currently occurring in the outer extent
of the cloud via triggering. 

As mentioned earlier, our identification is biased towards longer ETLSs that are visible as thin long structures pointing towards
massive stars from the base of the parental cloud or bubble wall in the 24 $\micron$ maps above background emission. Therefore, our
catalog lacks small-scale structures such as globulettes which are very small and somewhat round dark clouds \citep{gah07} and 
isolated knots that are detached from the parental clouds.
It is worth noting that using narrow-band H$\alpha$ observations taken with 2.6m Nordic Optical Telescope, La Palma, 
\citet{car03} \& \citet{gah07} have found $\sim$ 18 globulettes within the circular regions shown in Fig. 8. They investigated the position,
dimension, orientation of the globulettes, and found that most objects were smaller than 9 $kAU$ ($\sim$ 0.04 pc) and nearly 
round in shape. Many of them were associated with dense cores which might collapse to form planetary-mass objects. 

As we find only one star or at most a few stars at the tip of each ETLS,
 the scale of star formation in each cloud associated with ETLSs is very small. However, 
the total product can be considerable because a large number of such structures 
can be formed in H{\sc ii} region environment \citep[e.g., see simulation results of][]{wal13}.
We note  that recent studies based on the Herschel observations suggest that filaments and filament 
structures of lengths similar to that of ETLSs are ubiquitous in star forming clouds 
\citep[e.g.][]{arz11,mali12,palm13,andre14}. Numerical works show that filaments are prone to fragmentation due to their geometry where 
small-scale density perturbations have time to collapse locally before global collapse occurs \citep{pon11}. 
Therefore, it is difficult to conclude whether these aggregates are really due to triggering, or they are 
simply the sources of the filaments uncovered by the expanding H{\sc ii}
 region. However, the circumstantial evidence favors the former. For example, in general it has been found that young protostars or cores are distributed along the long axis of the filament more or less like beads strung on a wire \citep{tafa15,anat15,sam15}. 
In contrast, we find a single or at best a few  protostars only at the apex of the ETLSs facing the massive cluster; 
indicating their formation is very likely influenced by the comprehensive effect of expanding H{\sc ii} regions on the cold materiel of the ETLSs. 
It is difficult to conclude whether this effect is weak (i.e., temporarily increasing the star formation rate by inducing stars to form earlier) or strong (i.e., increasing the star formation efficiency by causing the birth of stars that would not otherwise form). 
High resolution sensitive observations of the W4 complex at infrared and millimeter bands, along with proper-motion 
and age measurements of the stars would reveal the prevalence of such structures and scale of star-formation
within them. Nonetheless, taking the results of the present work at face value it can be argued that star formation in small-scale structures such as pillars or elephant
trunks or globulettes is likely an important ingredient for the formation of low-mass stars at the outskirts of 
star-forming complexes. Therefore, it could be one of the possible viable mechanisms for forming 
isolated low-mass stars in molecular clouds.  
\begin{table}
\center
\caption{Location and nature of the sources at the tips of ETLSs in the W4 complex.}
\begin{tabular}{ccccc} \hline
S. No. &RA (J2000)& Dec (J2000)& Evolutionary stage\\
\hline
1  & 38.379875 & 60.423753 & Class~I \\ 
2  & 37.973150 & 60.530311 & Class~II\\
3  & 39.010692 & 60.611611 &  Class~II \\
4  & 37.612908 & 60.733211 &  Class~II \\
5  & 37.250650 & 61.430108 & Class~I \\  
6  & 38.274375 & 61.076200 &  Class~II\\
7  & 39.410421 & 61.086119 &  Class~II \\
8  & 38.448054 & 61.120731 &  Class~I\\
9  & 38.451850 & 61.112100 &  Class~II\\
10 & 37.758258 & 61.132481 &  Class~II\\
11 & 38.795842 & 61.133550 &  Class~II\\
12 & 37.173542 & 61.151639 &  Class~II\\
13 & 38.691429 & 61.192431 &  Class~II\\
14 & 39.380979 & 61.220169 &  Class~II\\
15 & 39.466100 & 61.226253 &  Class~II\\  
16 & 37.256721 & 61.258931 & Class~II\\
17 & 37.261350 & 61.257700 & Class~II\\
18 & 39.138429 & 61.317725 & Class~I\\
19 & 38.747588 & 61.333089 & Class~II\\ 
20 & 39.083142 & 61.395622 & Class~II\\
21 & 39.295542 & 61.435703 & Class~I\\
22 & 37.306521 & 61.296750 & Class~I\\
23 & 38.832171 & 60.434014 & Class~I\\
24 & 37.618654 & 60.553192 & Class~I\\
25 & 38.790225 & 61.329911 & Class~I\\
26 & 39.671125 & 61.860408 & Class~I\\
27 & 38.690279 & 61.190503 & Class~I\\
28 & 39.696529 & 61.860408 & Class~I\\
29 & 37.088583 & 61.266072 & Class~I\\
30 & 37.242608 & 61.432064 & Class~I\\
31 & 38.275258 & 61.066281 & Class~II\\
32 & 39.655446 & 61.849247 & Class~II\\
33 & 37.565021 & 61.696329 & Class~I\\
34 & 36.338496 & 60.966117 & Class~I\\
35 & 38.460296 & 60.365197 & Class~I\\
36 & 39.927787 & 62.012761 & Class~I\\
\hline    
\end{tabular}
\end{table}

\section{Conclusions}
In order to study the clustered and distributed mode of star formation in the W4 H{\sc ii} region,
we identified young stellar aggregates using homogeneous datasets from 2MASS, $Spitzer$ and WISE observations.
 We searched for signatures of star formation in the surroundings and 
investigated a probable scenario of triggered star formation at the border of the nebula. The spatial density distribution of the YSOs revealed three new clusterings/aggregates 
in the region, near BRC~7, CG~7S and south-east periphery of the H{\sc ii} region. 
Most of the identified YSOs in these three clusterings are low-mass stars with ages $\le$ 1 Myr.  
We also noticed many ETLSs in the H{\sc ii} region and most of them have young stars at their tips.  
We suggest that these may be the product of the interaction of the I/S-fronts with the dynamically evolving clumpy molecular cloud  
and that distributed low-mass young stars found at the outskirts of the large-complexes may be outcomes of feedback effects of massive stars. 

\acknowledgments
We thank the anonymous referee for the thorough reading of the manuscript and providing useful comments to improve it. 
NP acknowledges the financial support from the Department of Science \& Technology,
INDIA, through INSPIRE faculty award~IFA-PH-36. This publication makes use of data from the Two Micron All Sky Survey (a 
joint project of the University of Massachusetts and the Infrared Processing 
and Analysis Center/ California Institute of Technology, funded by the 
National Aeronautics and Space Administration and the National Science 
Foundation), archival data obtained with the {\it Spitzer Space Telescope} and {\it Wide Infrared Survey Explorer} 
(operated by the Jet Propulsion Laboratory, California Institute 
of Technology, under contract with the NASA.

\bibliography{r}
\listofchanges
\end{document}